\documentclass[aps,prl,twocolumn,superscriptaddress,floatfix]{revtex4-2}

\usepackage[T1]{fontenc}
\usepackage{lmodern}
\ifdefined\pdfgentounicode\input{glyphtounicode}\pdfgentounicode=1\fi
\usepackage{amsmath,amssymb,amsthm}
\usepackage{graphicx}
\usepackage{bm}
\usepackage{array}[=2016-10-06]
\usepackage{xcolor}
\definecolor{linkblue}{HTML}{1F5FBF}
\definecolor{citemagenta}{HTML}{C0157D}
\usepackage[colorlinks=true,linkcolor=linkblue,citecolor=citemagenta,urlcolor=linkblue]{hyperref}
\usepackage{orcidlink}
\hypersetup{pdftitle={The Fundamental Memory Ceiling of Noisy Quantum Reservoirs},pdfauthor={Markus Baumann, Jonas Stein, Claudia Linnhoff-Popien}}

\theoremstyle{plain}
\newtheorem{theorem}{Theorem}
\newtheorem{lemma}{Lemma}
\newtheorem{proposition}[lemma]{Proposition}
\newtheorem{corollary}[lemma]{Corollary}

\newcommand{\tr}{\operatorname{Tr}}
\newcommand{\ket}[1]{|#1\rangle}
\newcommand{\ketbra}[2]{|#1\rangle\!\langle #2|}
\newcommand{\nrm}[2]{\lvert #1\rvert_{#2}}
\newcommand{\Norm}[1]{\lVert #1\rVert}
\newcommand{\cD}{\mathcal D}
\newcommand{\cE}{\mathcal E}
\newcommand{\Nq}{\mathcal N_q}
\newcommand{\AD}{\mathcal{A\hspace{-0.6pt}D}}
\newcommand{\runin}[1]{\textit{\textbf{#1}}---}

\begin{document}

\let\SMaddcontentsline\addcontentsline
\renewcommand{\addcontentsline}[3]{}

\title{The Fundamental Memory Ceiling of Noisy Quantum Reservoirs}

\author{Markus Baumann\,\orcidlink{0009-0007-3575-1006}}
\email[Contact author: ]{markus.baumann@campus.lmu.de}
\author{Jonas Stein\,\orcidlink{0000-0001-5727-9151}}
\author{Claudia Linnhoff-Popien\,\orcidlink{0000-0001-6284-9286}}
\affiliation{Quantum Applications and Research Laboratory (QAR-Lab),
Ludwig-Maximilians-Universit\"at M\"unchen, Oettingenstra\ss e 67, 80538 M\"unchen, Germany}

\date{October 19, 2026}

\begin{abstract}
Fresh qubits, not size, set how long a noisy quantum system can remember. With $k$ input qubits reinitialized per time step, the longest memory any reservoir can attain under depolarizing noise is $\exp[\Theta(k)]$ time steps, whatever its size, dynamics, and input encoding. A universal upper bound and an explicit reservoir that attains the exponential scaling establish this optimal resource law. The bound depends on the dynamics only through the entropy that fresh qubits remove per time step, so it holds for any Markovian noisy dynamics without feedback, including measurement-free error correction.
\end{abstract}

\maketitle

\begin{figure}[t]
\centering
\includegraphics{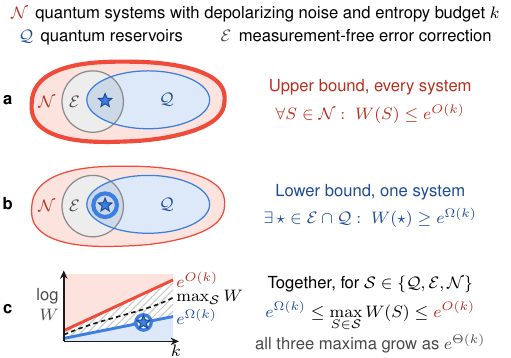}
\caption{\textbf{Scope of the memory limit} (schematic).
(a)~The upper bound of Theorem~\ref{thm:main} holds for every system in $\mathcal N$: dynamics with Markovian depolarizing noise, $k$ fresh qubits per time step, and no classical feedback, whatever its size and gate set.
(b)~One explicit system (star) in $\mathcal E\cap\mathcal Q$, the error-correcting reservoir of Fig.~\ref{fig:attain}(a), attains the lower bound, so it bounds the largest memory in both sets from below.
(c)~Since $\mathcal Q,\mathcal E\subseteq\mathcal N$, all three maxima grow as $e^{\Theta(k)}$ at fixed noise and accuracy [Fig.~\ref{fig:limit}(a)].}
\label{fig:scope}
\end{figure}

\runin{Introduction.}Reservoir computing processes time series by injecting inputs into a fixed physical system and training only a linear readout~\cite{Jaeger2001,Maass2002,JaegerHaas2004,Lukosevicius2009,Tanaka2019}. Quantum reservoir computing (QRC) uses interacting qubits for this purpose~\cite{Fujii2017,Nakajima2019,Mujal2021} and has been realized on superconducting processors with up to 120 qubits~\cite{Chen2020,Suzuki2022,Yasuda2023,Hu2024,Monzani2024}. For temporal tasks, memory is the decisive resource. A reservoir can use an input only as long as that input still influences its state, and on noisy hardware this influence fades at every time step. Can a noisy quantum reservoir, or any noisy quantum system supplied with fresh qubits [Fig.~\ref{fig:scope}], remember arbitrarily far back, provided it is made large enough?

\begin{figure*}[t]
\centering
\includegraphics{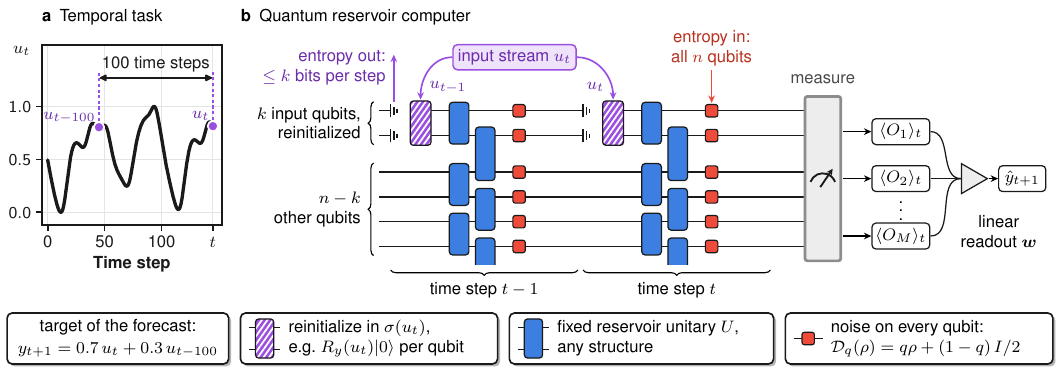}
\caption{\textbf{Noisy quantum reservoir computing.}
(a)~Forecasting $y_{t+1}=0.7\,u_t+0.3\,u_{t-100}$ needs a memory of 100 time steps.
(b)~Every time step reinitializes $k$ input qubits in $\sigma(u_t)$, applies a fixed unitary $U$ to all $n$ qubits, and depolarizes every qubit; a linear readout predicts $\hat y_{t+1}$. Noise injects entropy into all $n$ qubits, but only the $k$ input qubits carry it out.}
\label{fig:overview}
\end{figure*}

The answer is not obvious. A larger reservoir can spread information about past inputs over more degrees of freedom, which promises robustness to noise. Refreshing matters too: QRC discards and reinitializes the input qubits at every time step, which removes accumulated entropy [Fig.~\ref{fig:overview}(b)]~\cite{Fujii2017,Yasuda2023,Hu2024}, while noise brings entropy into every qubit at every time step.

Fading memory is well studied: echo-state and fading-memory criteria guarantee that old inputs eventually lose their influence~\cite{Jaeger2001,Boyd1985,MartinezPena2023,MorguiSancho2026}, and Markovian reservoirs forget exponentially fast~\cite{Sannia2026}. Dissipation and resets are recognized resources for QRC~\cite{Chen2020,Kubota2023,Sannia2024,Monzani2024,Freiheit2026}, and in noisy quantum computation fresh qubits are indispensable, since without them depolarizing noise permits only logarithmically deep computations~\cite{Aharonov1996,BenOr2013,StilckFranca2021,Mele2026}. But how much memory a given supply of fresh qubits can buy has remained open.

\begin{figure*}[t]
\centering
\includegraphics{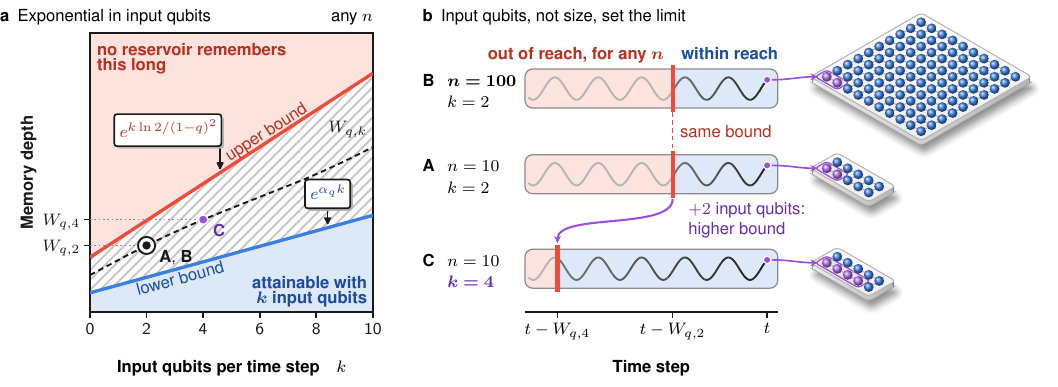}
\caption{\textbf{The memory limit} (schematic).
(a)~The optimum over reservoir sizes lies between two exponentials in $k$ (Theorem~\ref{thm:main}); A, B, and C mark the reservoirs of (b).
(b)~One input stream drives three reservoirs: A and B differ tenfold in size but share $k=2$ input qubits, and C is A with two more input qubits. Inputs older than the bound (red) are out of reach for every reservoir with the same $k$, so A and B share the bound $W_{q,2}$, and C's bound $W_{q,4}$ lies further back.}
\label{fig:limit}
\end{figure*}

The memory ceiling is set by fresh qubits, independently of size. Let $k$ input qubits be reinitialized at each time step, and let every qubit undergo depolarizing noise that keeps its state with probability $q$. The memory depth of a reservoir is the number of past inputs that still change its state by more than $\varepsilon$. The largest memory depth that any such reservoir can attain grows exponentially with $k$ and does not depend on the total number of qubits $n$ [Fig.~\ref{fig:limit}(a)]:
\begin{equation}
W_{q,k}(\varepsilon)=\exp\!\big[\Theta_q(k)\big]\quad\text{at fixed }\varepsilon.
\label{eq:main}
\end{equation}
Equation~\eqref{eq:main} turns size into supply: retaining an input for $T$ time steps needs $\Theta_q(\log T)$ reinitialized qubits per time step, and that many suffice [Fig.~\ref{fig:limit}(b)]. The sufficiency has been known since von Neumann, through the repetition code with majority vote~\cite{vonNeumann1956}. New is the necessity within this noise and entropy-budget model: no system, however large, however entangled, and whatever its unitary, gate set, or input encoding, keeps an input for $T$ time steps with fewer than $\Omega_q(\log T)$ fresh qubits per time step. Counting entropy does not give this, because even a highly mixed state can store a bit perfectly. The proof instead works with purity and concentration: echo states have a purity bounded independently of $n$; a state difference resists noise only through concentration, which at fixed distinguishability costs purity. The purity cap therefore forces exponential forgetting. Earlier bounds constrain total qubits~\cite{Fawzi2022} or Clifford circuits~\cite{Nelson2026}, rather than the fresh-qubit rate~\cite[Sec.~VIII\,B]{SM}.

This budget formulation makes the upper bound independent of architecture: it covers Markovian dynamics with a fixed local depolarizing layer, provided every entropy sink is charged to the budget. Without classical feedback, it covers input-driven variational quantum recurrent neural networks~\cite{Takaki2021}, repeated-interaction models with fresh ancillas~\cite{Cattaneo2021}, and dynamical circuits with resets~\cite{Yasuda2023,Hu2024}, including measurement-free error correction~\cite{Heussen2024}.

\runin{A reservoir with fresh qubits.}A reservoir of $n$ qubits receives one classical input $u_t$ per time step $t$ [Fig.~\ref{fig:overview}(a),(b)]. At each step, the $k$ input qubits are discarded and reinitialized in the encoding state $\sigma(u_t)$. A fixed reservoir unitary $U$ then acts on all qubits. Finally, every qubit undergoes depolarizing noise $\cD_q(\rho)=q\rho+(1-q)\tr(\rho)I/2$. Altogether,
\begin{equation}
\rho_t=\cD_q^{\otimes n}\Bigl(U\bigl[\sigma(u_t)\otimes\tr_{\mathrm{in}}\rho_{t-1}\bigr]U^\dagger\Bigr),
\label{eq:model}
\end{equation}
where $\tr_{\mathrm{in}}$ is the partial trace over the input register. A trained linear readout turns the state into a prediction $\hat y_{t+1}=\sum_iw_i\tr(O_i\rho_t)=\tr(F\rho_t)$. This covers the original QRC proposal~\cite{Fujii2017} and protocols that measure and reset qubits without feedback~\cite{Yasuda2023,Hu2024}. We impose no restriction on $\sigma(u)$ or $U$; the unitary may even depend on the input.

The input qubits enter our analysis through a single number. A channel $\cE$ with $\cE(I)\le2^bI$, where $I$ is the identity, can lower the entropy of the completely mixed state by at most $b$ bits; we call $b$ its entropy \emph{budget}. Because $\sigma(u)\le I$, the injection step $\cE_u(X)=U[\sigma(u)\otimes\tr_{\mathrm{in}}X]U^\dagger$ has budget $k$.

Noise erases the initial state, so the reservoir eventually reaches a state that depends only on the input history $h=(\dots,u_{t-1},u_t)$, its \emph{echo state} $\rho(h)$~\citetext{\citealp{Jaeger2001,MartinezPena2023}; \citealp[Sec.~III\,F]{SM}}. The \emph{memory depth} $W(\varepsilon)$ is the smallest number $L$ of time steps such that any two input histories that agree on their last $L$ inputs lead to echo states with $\tfrac12\Norm{\rho(h)-\rho(h')}_1\le\varepsilon$. Inputs older than $W(\varepsilon)$ change every linear readout by at most $2\varepsilon\Norm F$~\cite[Sec.~III\,F]{SM}, so the memory depth, unlike the memory capacity~\cite{Jaeger2002}, limits all readouts at once. We write $W_{q,k}(\varepsilon)$ for the largest memory depth of any reservoir with $k$ input qubits, maximized over $n$, $U$, and $\sigma$.

\runin{The memory limit.}Our main result bounds $W_{q,k}$ from both sides.

\begin{theorem}[Memory limit]\label{thm:main}
Let $0<q<1$.\\[2pt]
(i)~For $0<\varepsilon<1$, every reservoir with $k$ input qubits, whatever its number of qubits, reservoir unitary, and input encoding, has
\begin{equation}
W(\varepsilon)\le\Bigl\lceil\delta^{-1}\ln\frac1\varepsilon\Bigr\rceil,\qquad\delta^{-1}=2^{(1+4q)/(1-q)}\,2^{\beta_qk},
\label{eq:upper}
\end{equation}
with $\beta_q=q(1+q^2)/[(1-q)(1-q^2)]$, and \mbox{$\delta^{-1}\le2^{k/(1-q)^2}$} for $k\ge4$.\\[2pt]
(ii)~For $0<\varepsilon<q$ and every $k$ with $e^{\alpha_q(k-3)}\ge4$, there is a reservoir with at most $2k$ qubits that has
\begin{equation}
W(\varepsilon)\ge\tfrac14\,e^{\alpha_q(k-3)}\ln\frac q\varepsilon ,
\label{eq:lower}
\end{equation}
where $\alpha_q=\frac12\ln\frac1{1-q^2}$.
\end{theorem}

At fixed noise strength and accuracy, both bounds are exponential in $k$ and independent of $n$: the fresh-qubit rate sets the optimal memory scaling. One explicit reservoir attains the lower bound [Fig.~\ref{fig:scope}], while the upper bound covers every reservoir; neither size nor entanglement nor the choice of unitary can change this scaling class. The two exponents differ, so the theorem fixes the class but not the exact rate~\cite[Sec.~III\,G]{SM}. For readouts of finite precision, the upper bound carries over to the memory capacity~\citetext{\citealp[Sec.~III\,F]{SM}; \citealp{Jaeger2002}}.

The upper bound uses the reservoir only through its budget: any channel with budget $k$ may replace the injection step, resets and measurement-free recovery operations included~\cite[Sec.~VI]{SM}. The theorem concerns echo states: a larger reservoir needs a longer washout of its initial state, $\Theta_q(\log n)$ time steps for $n$ redundant copies of a bit~\cite[Sec.~IV]{SM}, without raising the ceiling for driven memory.

\runin{Why no reservoir can remember longer.}A proof must defeat every strategy a reservoir might use to hide information: spreading it over many qubits, entangling it, or using its input qubits for repair. Ours follows the difference between two reservoir states with different input histories and asks how well the states can be distinguished and how concentrated their difference is. Two properties of depolarizing noise~\cite{King2014,Cubitt2015} answer both and yield the bound in three steps.

\emph{Step 1: Noise caps the purity.} We measure the purity of a state by $\log_2P^2=n-S_2(\rho)$, the number of pure qubits' worth of purity it holds, where $S_2$ is the R\'enyi-2 entropy. Depolarizing noise damps the many-qubit correlations of which purity is made; this is hypercontractivity~\cite{King2014}. As a result, every time step multiplies the log-purity by $\theta=2q^2/(1+q^2)<1$, while the input qubits add at most $\theta k$. In a reservoir of any size, every echo state therefore holds at most $\log_2P_*^2=2kq^2/(1-q^2)$ pure qubits' worth, about 8.5 per input qubit at $q=0.9$~\cite[Sec.~III\,B]{SM}.

\emph{Step 2: Spread-out differences are erased.} Low purity alone does not mean forgetting: $\ketbra00\otimes I/2^{n-1}$ and $\ketbra11\otimes I/2^{n-1}$ are almost completely mixed, yet perfectly distinguishable. Their difference, however, is spread over the whole Hilbert space, and one layer of noise erases the fraction $1-q$ of their distinguishability. Reverse hypercontractivity makes this precise~\cite{Cubitt2015}: noise forces two states to overlap unless their difference is concentrated, so one layer of noise erases a fixed fraction of every spread-out difference.

\emph{Step 3: Concentration costs purity.} A distinguishable difference concentrated on a fraction $2^{-m}$ of the Hilbert space, such as a bit stored in $m$ redundant copies, requires about $m$ pure qubits' worth of purity, and Step~1 caps the purity. A distance $V$ that weighs distinguishability against concentration therefore shrinks at every time step $\Phi_u$, for any two reservoir states $\rho,\rho'$ and any input $u$:
\begin{equation}
V\bigl(\Phi_u\rho-\Phi_u\rho'\bigr)\le(1-\delta)\,V(\rho-\rho'),
\label{eq:contraction}
\end{equation}
with the forgetting rate $\delta$ of Eq.~\eqref{eq:upper}~\cite[Secs.~III\,D--E]{SM}. Whatever the reservoir does, every time step erases at least the fraction $\delta$ of its past. After $L$ time steps, two echo states are therefore within trace distance $e^{-\delta L}$; this is Theorem~\ref{thm:main}(i).

Without fresh qubits ($k=0$), as for inputs encoded in input-dependent unitaries, the echo state is completely mixed for every input sequence, so such a reservoir retains nothing~\citetext{\citealp{Aharonov1996}; \citealp[Sec.~III\,F]{SM}}.

\begin{figure*}[t]
\centering
\includegraphics{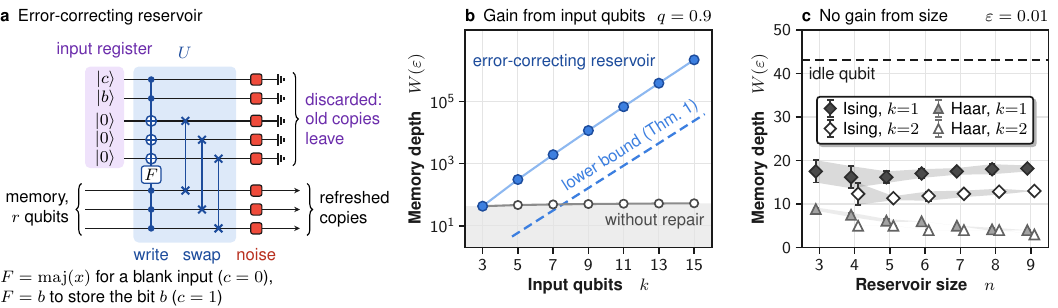}
\caption{\textbf{Attaining the limit.}
(a)~Error-correcting reservoir ($r=3$, $k=5$): $U$ writes the new bit, or the majority of the stored copies, into fresh ancillas and swaps them into the memory; the old copies leave with the input register.
(b)~With its fresh qubits used for repair, the memory depth of the error-correcting reservoir (exact single-qubit readout, $n=2k-2$) grows exponentially with $k$, above the lower bound of Theorem~\ref{thm:main}; without repair (gray), the same $r=k-2$ copies gain almost nothing. Depths are continuous threshold crossings, rounded up to integers.
(c)~Within the sampled ensembles, standard reservoirs gain nothing from size at $k=1$ (filled) or $k=2$ (open); mean $\pm$ s.d., six realizations per $k$. (b),~(c)~$q=0.9$, $\varepsilon=0.01$.}
\label{fig:attain}
\end{figure*}

\runin{A reservoir that attains the limit.}A single reservoir suffices to attain the optimal scaling of $W_{q,k}$; it may be classical because the upper bound covers every quantum reservoir. Our exactly solvable error-correcting reservoir repairs its memory with fresh qubits [Fig.~\ref{fig:attain}(a)]. It stores a bit in $r$ memory qubits as a repetition code and corrects it by majority vote~\cite{vonNeumann1956}. At every time step, its unitary computes the majority of the stored copies, writes it into $r$ freshly reinitialized ancillas, and swaps the ancillas into the memory. The old copies, which carry the noise, end up in the input register and are discarded at the next injection, so the entropy leaves with them. Two further input qubits tell the reservoir whether to keep its memory or to store a new bit, so that $k=r+2$~\cite[Sec.~V]{SM}. The unitary only permutes computational basis states: the construction is classical reversible logic run on qubits, and it is measurement-free error correction~\cite{Heussen2024} in its simplest form.

Because all states remain diagonal, every qubit behaves as a classical bit that flips with probability $(1-q)/2$ per time step, and the response can be computed exactly. The stored bit is lost only if most of its $r$ copies flip within one time step, which happens with probability $p_r\le(1-q^2)^{r/2}=e^{-\alpha_qr}$ by a Chernoff bound with the exact exponent~\cite[Sec.~V]{SM}. A bit $b$ stored $t$ time steps ago therefore leaves every memory qubit with
\begin{equation}
\langle Z\rangle_t=(-1)^b\,q\,(1-2p_r)^t .
\label{eq:response}
\end{equation}
The signal decays at a rate exponentially small in $r$. Two input histories that differ only in this bit agree on their last $t$ inputs, yet their echo states remain at trace distance at least $q(1-2p_r)^t$; with the bound on $p_r$, this gives Theorem~\ref{thm:main}(ii). At $q=0.9$ and $\varepsilon=0.01$, the single-qubit memory depth exceeds 300 time steps for $k=5$ and two million for $k=15$. For the odd values $k=5,\ldots,15$ in Fig.~\ref{fig:attain}(b), each pair of additional input qubits increases this depth by more than a factor of five. The same copies without majority vote keep the bit for at most 53 time steps; the fresh qubits extend memory because the reservoir uses them for repair.

\runin{Standard reservoirs operate far below the limit.}The simulations do not test the bound; they locate standard reservoirs relative to it. We simulated transverse-field Ising reservoirs with random all-to-all couplings, the workhorse of QRC~\cite{Fujii2017,Nakajima2019,MartinezPena2021}, and Haar-random unitaries at $q=0.9$, and estimated their memory depth $W$ at $\varepsilon=0.01$~\cite{SM}. Within these ensembles, size does not extend the memory [Fig.~\ref{fig:attain}(c)]: with one input qubit, the Ising reservoir remembers 16 to 18 time steps for every size from $n=3$ to $9$, and the Haar reservoir forgets faster as it grows ($W$ drops from 9 to 4). More input qubits raise the ceiling but, in generic dynamics, lower the memory, because reinitialization also erases what those qubits had stored: at $n=8$, raising $k$ from 1 to 4 lowers $W$ from 18 to 8 time steps (Ising) and from 4 to 3 (Haar), even when the additional qubits are reinitialized in $\ket0$~\cite{SM}. Both stay below an idle qubit that merely holds the input ($W=43$), whereas the error-correcting reservoir of the same size ($n=8$, $k=5$) remembers over 300 time steps. The difference lies in how the dynamics uses the fresh qubits.

\runin{Beyond reservoirs: an entropy budget.}The same bound covers other noise and refresh operations: with a fixed local depolarizing component, additional channels with total budget $b$ replace $k$ by $k+b$~\cite[Sec.~VI]{SM}. Pauli errors, dephasing, coherent errors, coherent crosstalk, and projective measurements whose outcomes only enter the readout~\cite{Mujal2023} are unital: their zero budget does not raise the upper bound. Every unital qubit channel with full-rank Choi matrix contains a depolarizing component, so the theorem covers generic independent unital noise~\cite[Sec.~VI]{SM}. Entropy-removing channels contribute a positive budget. Resetting $m$ qubits adds $m$ bits. Amplitude damping with probability $0<\gamma<1$ adds $n\log_2(1+\gamma)$ bits; relaxing idle qubits then act as a refrigerator~\cite{BenOr2013}, allowing a reservoir with two input qubits to remember for a time exponential in $n$~\citetext{\citealp[Sec.~VI]{SM}; \citealp{Hoeffding1963}}. With no additional entropy sink, a measurement-free recovery using $k$ fresh ancillas per round belongs, together with its data qubits, to $\mathcal N$ [Fig.~\ref{fig:scope}]. After initialization washout, it can protect a logical bit for at most $\exp[O_q(k)]$ rounds, regardless of size. Classical syndrome feedback is outside the model and can lift the ceiling~\cite[Sec.~VI]{SM}. Without a depolarizing component or with non-Markovian noise~\cite{Sannia2026}, memory can be unbounded~\cite[Sec.~VI]{SM}.

\runin{Conclusion.}After initialization washout, fresh-qubit supply sets a size-independent memory ceiling for Markovian quantum dynamics with a fixed local depolarizing component, no classical feedback, and no other entropy sink. Fresh qubits are necessary but insufficient: an explicit reservoir uses them for repair and attains exponential scaling, whereas standard reservoirs without repair operate far below the ceiling. The ceiling bounds how far back a task can reach; how well a reservoir processes the inputs within that range is a separate question~\cite{Cao2026}. Open problems include closing the exponent gap, whether a quantum code can beat the classical rate $\alpha_q$, what limits memory with feedback or non-Markovian noise, and testing the scaling by varying $k$ at fixed $n$ with mid-circuit resets~\cite{Yasuda2023,Hu2024}.

\begin{acknowledgments}
\runin{Acknowledgments.}Claude Opus 5.5 (Anthropic) assisted with manuscript preparation, checks of derivations and references, and simulation and plotting code. Markus Baumann reviewed the outputs and takes responsibility for the final content.
\end{acknowledgments}

\runin{Data availability.}Code and data that reproduce all numbers and figures are provided with the Supplemental Material~\cite{SM} and in the code repository~\cite{CodeRepository}.

\nocite{Xiong2025,Dao2026}
\bibliography{refs}

\clearpage
\onecolumngrid
\setcounter{equation}{0}
\setcounter{figure}{0}
\setcounter{table}{0}
\setcounter{section}{0}
\setcounter{lemma}{0}
\setcounter{page}{1}
\renewcommand{\theequation}{S\arabic{equation}}
\renewcommand{\theHequation}{S\arabic{equation}}
\renewcommand{\thefigure}{S\arabic{figure}}
\renewcommand{\thetable}{S\arabic{table}}
\renewcommand{\thelemma}{S\arabic{lemma}}
\renewcommand{\thepage}{S\arabic{page}}
\renewcommand{\theHfigure}{S\arabic{figure}}
\renewcommand{\theHtable}{S\arabic{table}}
\setcounter{secnumdepth}{3}

\begin{center}
{\large\bfseries Supplemental Material:\\[3pt] The Fundamental Memory Ceiling of Noisy Quantum Reservoirs}\\[10pt]
Markus Baumann\,\orcidlink{0009-0007-3575-1006}, Jonas Stein\,\orcidlink{0000-0001-5727-9151}, and Claudia Linnhoff-Popien\,\orcidlink{0000-0001-6284-9286}\\[3pt]
{\small\itshape Quantum Applications and Research Laboratory (QAR-Lab),\\
Ludwig-Maximilians-Universit\"at M\"unchen, Oettingenstra\ss e 67, 80538 M\"unchen, Germany}
\end{center}

This Supplemental Material contains the proofs of the main text, the exact analysis of the error-correcting reservoir, the entropy budget for noise beyond depolarizing noise, the details of the simulations, the relation to earlier limits on noisy computation, and a guide to the code and data. Section~\ref{sm:overview} explains the idea of the proof in words, and every later section can be read on its own. Equations, figures, tables, and lemmas carry the prefix S; numbers without prefix and references refer to the main text.

\let\addcontentsline\SMaddcontentsline
\tableofcontents
\vspace{8pt}

\section{How the proof works}
\label{sm:overview}

The upper bound of Theorem~\ref{thm:main} rests on a single statement: every time step shrinks the difference between any two reservoir states by a fixed fraction, for any reservoir dynamics. The difficulty is that neither of the two obvious ways to measure a difference shrinks on its own.
\begin{itemize}
\item The \emph{trace distance}, which says how well two states can be distinguished, can barely change in one time step: a bit stored in many redundant copies passes a layer of noise almost intact.
\item The \emph{2-norm}, which is large when a difference is concentrated on a small part of the Hilbert space, can even grow, because the fresh input qubits make the state purer.
\end{itemize}
The proof combines the two in the three steps of the main text.
\begin{enumerate}
\item \emph{Noise caps the purity} of every state (Lemma~\ref{lem:cap}).
\item \emph{One layer of noise erases a fixed fraction of every difference that is not concentrated} (Lemma~\ref{lem:forget}).
\item \emph{Concentration costs purity}, so a combination of trace distance and 2-norm shrinks at every time step (Proposition~\ref{prop:contraction}).
\end{enumerate}
Steps 1 and 2 each rest on one known inequality for depolarizing noise, hypercontractivity~\cite{King2014} and reverse hypercontractivity~\cite{Cubitt2015}; all other steps are elementary. Section~\ref{sm:memory} turns the shrinking distance into the memory bound of Theorem~\ref{thm:main}(i). Sections~\ref{sm:washoutlogn}--\ref{sm:numerics} derive the remaining statements of the main text: the washout of the initial state, the error-correcting reservoir and Theorem~\ref{thm:main}(ii), the entropy budget, and the simulations.

\section{What the proof keeps track of}
\label{sm:norms}

\subsection{Averaged norms}

For an operator $X$ on $n$ qubits, let $d=2^n$ and $\tau(X)=\tr(X)/d$, the average eigenvalue of $X$. We use
\begin{equation}
\nrm Xp=\bigl(\tau|X|^p\bigr)^{1/p},
\label{eq:S-norm}
\end{equation}
the $p$-th root of the average $p$-th power of the singular values of $X$ (the normalized Schatten $p$-norm). Averaging instead of summing makes $\nrm Ip=1$ for every $p$, so that no factors of $d$ appear in the proofs. We need $p=1$, $p=2$, and, in one place, a value of $p$ between $0$ and $1$, for which the same formula defines a useful quantity, although not a norm. We use three standard facts about such averages: (i) $\nrm Xp$ grows with $p$; (ii) $p\mapsto\ln\tau|X|^p$ is convex; and (iii) $\nrm Xp\le\nrm X1^{1-\theta}\nrm X2^{\theta}$ whenever $1/p=1-\theta/2$ (H\"older's inequality).

\subsection{Distinguishability, purity, and concentration}

The proof tracks three properties of the operator $X=d(\rho-\rho')$ for two reservoir states $\rho$ and $\rho'$.
\begin{itemize}
\item \emph{Distinguishability}: $\nrm X1=\Norm{\rho-\rho'}_1$ is twice the trace distance. The best measurement tells $\rho$ from $\rho'$ with success probability $\frac12+\frac14\nrm X1$.
\item \emph{Purity}: for a single state, $P=\nrm{d\rho}2=(d\tr\rho^2)^{1/2}$, and $\log_2P^2=n-S_2(\rho)$, where $S_2=-\log_2\tr\rho^2$ is the R\'enyi-2 entropy. It counts \emph{pure qubits' worth} of purity: $\log_2P^2$ equals $0$ for the completely mixed state, $n$ for a pure state, and $m$ for a state with $m$ pure and $n-m$ completely mixed qubits.
\item \emph{Concentration}: the ratio $\nrm X2/\nrm X1\ge1$. It equals one if all eigenvalues of $X$ have the same magnitude, and it is large if $X$ lives on a small subspace.
\end{itemize}
The following examples show how widely the concentration varies. The difference between $\ketbra00\otimes I/2^{n-1}$ and $\ketbra11\otimes I/2^{n-1}$ has concentration $1$: it is spread over the whole Hilbert space. A bit stored in $m$ redundant copies, $\ketbra00^{\otimes m}$ versus $\ketbra11^{\otimes m}$ with the other qubits completely mixed, has concentration $2^{(m-1)/2}$. Two orthogonal pure states have the largest possible concentration, $2^{(n-1)/2}$. In all three examples, a concentrated difference requires purity, and this holds in general: since $\nrm X2\le\nrm{d\rho}2+\nrm{d\rho'}2$, two states with purity at most $P$ have a difference with $\nrm X2\le2P$.

\subsection{The fresh qubits are the only source of purity}

\begin{lemma}[Budget]\label{lem:norms}
The injection step $\cE_u(X)=U[\sigma(u)\otimes\tr_{\mathrm{in}}X]U^\dagger$ obeys, for every Hermitian $X$,
\begin{equation}
\nrm{\cE_uX}1\le\nrm X1,\qquad\nrm{\cE_uX}2\le2^{k/2}\,\nrm X2 .
\label{eq:S-norms}
\end{equation}
\end{lemma}

\begin{proof}
The first inequality says that no physical process makes two states more distinguishable (contractivity of the trace norm). For the second, write $X=\sum_{i,j}\ketbra ij\otimes X_{ij}$, where $\ket i$ runs over the $2^k$ basis states of the input register. Then $\tr_{\mathrm{in}}X=\sum_iX_{ii}$, and the Cauchy--Schwarz inequality gives
\[
\tr(\tr_{\mathrm{in}}X)^2\le2^k\sum_i\tr X_{ii}^2\le2^k\tr X^2 .
\]
Since $\tr\sigma(u)^2\le1$ and $U$ does not change traces,
\[
\tr(\cE_uX)^2=\tr\sigma(u)^2\,\tr(\tr_{\mathrm{in}}X)^2\le2^k\tr X^2 .
\]
Dividing by $d$ gives the claim.
\end{proof}

Discarding $k$ qubits thus raises the 2-norm of an operator by at most a factor $2^{k/2}$. For a state, $X=d\rho$, a time step adds at most $k$ pure qubits' worth of purity, so the fresh qubits are the only source of purity.

The same two bounds hold, with $k$ replaced by $b$, for every channel $\cE$ with $\cE(I)\le2^bI$; for the 2-norm, this follows from the Kadison--Schwarz inequality $(\cE X)^2\le\Norm{\cE(I)}\,\cE(X^2)$. We call $b$ the \emph{budget} of $\cE$, as in the main text. The injection step has $b=k$, because $\cE_u(I)=2^kU[\sigma(u)\otimes I]U^\dagger\le2^kI$; unitaries and all unital channels have $b=0$; and a reset of $m$ qubits has $b=m$ (Sec.~\ref{sm:budget}). In the following, $\cE$ denotes any channel with a budget of $b$ bits, and $b=k$ for the reservoirs of the main text.

\section{Proof of the upper bound}
\label{sm:upper}

Throughout this section, $\Nq=\cD_q^{\otimes n}$ denotes one layer of depolarizing noise, so that one time step is $\Phi_u=\Nq\circ\cE_u$.

\subsection{Two properties of depolarizing noise}
\label{sm:ineq}

\paragraph*{Noise smooths concentrated operators (hypercontractivity).} For every Hermitian $Y$,
\begin{equation}
\nrm{\Nq Y}2\le\nrm Y{1+q^2} .
\label{eq:S-king}
\end{equation}
This is King's hypercontractivity theorem for products of unital qubit channels~\cite[Corollary~2]{King2014}, which gives $\nrm{\Nq Y}r\le\nrm Yp$ whenever $q^2\le(p-1)/(r-1)$; here $r=2$ and $p=1+q^2$. Because $1+q^2<2$, the norm on the right is weaker than the one on the left, so noise turns operators with a bounded $(1+q^2)$-norm into operators with a bounded 2-norm. The mechanism is the damping of correlations: written as a sum of Pauli strings, an operator loses the factor $q$ for every qubit on which a string acts, and concentrated operators, such as projectors onto a few states, consist mostly of long strings. As a single-qubit check, $Y=2\ketbra00=I+Z$ becomes $I+qZ$, and Eq.~\eqref{eq:S-king} reads $1.345\le1.364$ at $q=0.9$. The extension from positive to Hermitian $Y=Y_+-Y_-$ is immediate. Since $\Nq$ is positive and self-adjoint with respect to $\tau$,
\[
\nrm{\Nq Y}2^2=\tau\bigl[Y\,\Nq^2(Y)\bigr]\le\tau\bigl[|Y|\,\Nq^2(|Y|)\bigr]=\nrm{\Nq|Y|}2^2,
\]
while $\nrm{|Y|}p=\nrm Yp$.

\paragraph*{Noise forces overlap (reverse hypercontractivity).} For positive operators $G_1,G_2\ge0$,
\begin{equation}
\tau\bigl[G_1\,\Nq(G_2)\bigr]\ge\nrm{G_1}{1-q}\,\nrm{G_2}{1-q} .
\label{eq:S-cubitt}
\end{equation}
This is the two-point reverse hypercontractive inequality of Cubitt \emph{et al.}~\cite[Corollary~14]{Cubitt2015}, which holds for all exponents $\bar p_1,\bar p_2<1$ with $q\le\sqrt{(1-\bar p_1)(1-\bar p_2)}$; we use the symmetric choice $\bar p_1=\bar p_2=1-q$. To interpret it, normalize $\tau G=1$, so that $G$ is $d$ times a density matrix. Then $\nrm G{1-q}$ lies between $0$ and $1$: it equals one for $G=I$ and is small only if $G$ is concentrated on a small subspace. After one layer of noise, two positive operators therefore overlap substantially unless they are concentrated. For example, the scaled orthogonal projectors $2\ketbra00$ and $2\ketbra11$ of a single qubit overlap by $1-q$ after one layer of noise.

\subsection{Step 1: Noise caps the purity}
\label{sm:cap}

\begin{lemma}[Purity cap]\label{lem:cap}
Let $\theta=2q^2/(1+q^2)$. For every Hermitian $X$,
\begin{equation}
\nrm{\Nq\cE X}2\le2^{b\theta/2}\,\nrm X1^{1-\theta}\,\nrm X2^{\theta}.
\label{eq:S-cap}
\end{equation}
\end{lemma}

\begin{proof}
Hypercontractivity~\eqref{eq:S-king}, fact (iii) with $p=1+q^2$ (which fixes $\theta$), and Lemma~\ref{lem:norms} give
\[
\nrm{\Nq\cE X}2\le\nrm{\cE X}{1+q^2}\le\nrm{\cE X}1^{1-\theta}\,\nrm{\cE X}2^{\theta}\le\nrm X1^{1-\theta}\bigl(2^{b/2}\nrm X2\bigr)^{\theta}. \qedhere
\]
\end{proof}

\begin{figure}[t]
\centering
\includegraphics{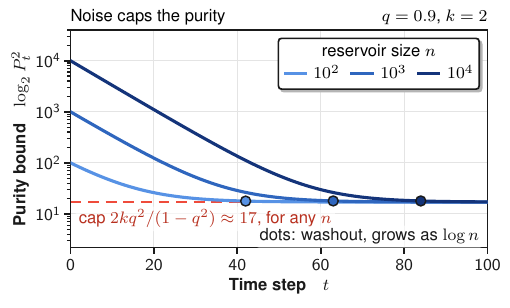}
\caption{\textbf{Noise caps the purity.} Purity bound $\log_2P_t^2$ obtained by iterating Eq.~\eqref{eq:S-purityrec} with equality, for reservoirs of $n$ qubits that start in a pure state, at $q=0.9$ and $b=k=2$. Every bound approaches the same cap $\log_2P_*^2=2kq^2/(1-q^2)$ (dashed) for every size; only the washout time $t_0$ of Sec.~\ref{sm:memory} (dots) grows, as $\log n$.}
\label{figS:purity}
\end{figure}

After the noise, the 2-norm is thus controlled by a weighted geometric mean of the 1-norm and the 2-norm before it. For a state, $X=d\rho$ has $\nrm X1=1$, so the purity obeys $P_t\le2^{b\theta/2}P_{t-1}^{\,\theta}$, or
\begin{equation}
\log_2P_t\le\theta\log_2P_{t-1}+\tfrac12b\theta .
\label{eq:S-purityrec}
\end{equation}
The log-purity is multiplied by $\theta<1$ at every time step, and the budget adds at most $b\theta/2$. Its fixed point is the \emph{purity cap}
\begin{equation}
\log_2P_*=\frac{bq^2}{1-q^2},
\label{eq:S-Pstar}
\end{equation}
which is the purity cap of the main text for $b=k$. The purity approaches the same cap for every size of the reservoir; only the time this takes grows, as $\log n$ [Fig.~\ref{figS:purity} and Sec.~\ref{sm:memory}].

In the proofs below, we use the cap in the form $2^{b\theta/2}=P_*^{\,\bar\theta}$ with $\bar\theta=1-\theta=(1-q^2)/(1+q^2)$.

\subsection{Step 2: Spread-out differences are erased}
\label{sm:forget}

\begin{lemma}[Forced forgetting]\label{lem:forget}
Let $a=2q/(1-q)$ and $c_q=2^{-1/(1-q)}$. Every traceless Hermitian $Y\ne0$ with concentration $K=\nrm Y2/\nrm Y1$ obeys
\begin{equation}
\nrm{\Nq Y}1\le\nrm Y1\bigl[1-c_q\,(2K)^{-a}\bigr].
\label{eq:S-forget}
\end{equation}
\end{lemma}

One layer of noise thus removes at least the fraction $c_q(2K)^{-a}$ of the distinguishability. This fraction is small only if the difference is concentrated.

\begin{proof}
\emph{Split the difference.} Write $Y=Y_+-Y_-$ with $Y_\pm\ge0$ supported on orthogonal subspaces. Because $Y$ is traceless, both parts have the same weight, $\tau Y_\pm=\frac12\nrm Y1$, so $G_\pm=2Y_\pm/\nrm Y1$ are $d$ times density matrices. Their purity is limited by the concentration of $Y$: $\nrm{G_\pm}2\le2K$.

\emph{Distinguishability is the success of the best test.} A measurement with outcomes $M$ and $I-M$ that guesses ``$+$'' on $M$ confuses the noisy parts with error probabilities that sum to $\mathrm{err}(M)=\tau[(I-M)\Nq G_+]+\tau[M\Nq G_-]$. The trace norm is attained by the best such test, which gives
\begin{equation}
\nrm{\Nq Y}1=\nrm Y1\Bigl[1-\min_{0\le M\le I}\mathrm{err}(M)\Bigr].
\label{eq:S-testerror}
\end{equation}

\emph{Every test errs.} Either $M$ or $I-M$ accepts at least half of the Hilbert space; say $\tau M\ge\frac12$. Reverse hypercontractivity~\eqref{eq:S-cubitt} then bounds the probability that $M$ accepts the noisy negative part,
\[
\mathrm{err}(M)\ge\tau[M\Nq G_-]\ge\nrm M{1-q}\,\nrm{G_-}{1-q},
\]
and both factors are bounded from below. A test that accepts half of the Hilbert space is not concentrated: $M^{1-q}\ge M$ gives $\nrm M{1-q}\ge(\tau M)^{1/(1-q)}\ge c_q$. A state part with limited purity is not concentrated either. By fact (ii), $\ln\tau G_-^p$ is convex in $p$, and $\tau G_-=1$ at $p=1$, which lies between $1-q$ and $2$; hence
\[
\tau G_-^{1-q}\ge\bigl(\tau G_-^2\bigr)^{-q},\quad\text{that is,}\quad\nrm{G_-}{1-q}\ge\nrm{G_-}2^{-a}\ge(2K)^{-a}.
\]
Therefore $\mathrm{err}(M)\ge c_q(2K)^{-a}$ for every test, and Eq.~\eqref{eq:S-testerror} proves the lemma.
\end{proof}

\subsection{Step 3: One distance shrinks at every time step}
\label{sm:contract}

Lemma~\ref{lem:forget} needs a bound on the concentration, and Lemma~\ref{lem:cap} provides one. We therefore measure a difference by both norms at once,
\begin{equation}
V(X)=\max\Bigl\{\nrm X1,\ \frac{\nrm X2}{2P_*}\Bigr\}.
\label{eq:S-V}
\end{equation}
That is, $V$ is twice the trace distance, unless the difference is so concentrated that its 2-norm, measured in units of twice the purity cap, is larger. For the difference of two states with purity at most $P_*$, both entries are at most 2 (Sec.~\ref{sm:norms}).

\begin{proposition}[Contraction]\label{prop:contraction}
For every traceless Hermitian $X$,
\begin{equation}
V(\Nq\cE X)\le(1-\delta)\,V(X),\qquad\delta=c_q\bigl(4\cdot2^{b/2}P_*\bigr)^{-a}.
\label{eq:S-contraction}
\end{equation}
\end{proposition}

\begin{proof}
Both sides grow linearly with $X$, so it suffices to take $V(X)=1$, that is, $\nrm X1\le1$ and $\nrm X2\le2P_*$.

\emph{The 2-norm part shrinks.} Lemma~\ref{lem:cap} and $2^{b\theta/2}=P_*^{\,\bar\theta}$ give
\[
\nrm{\Nq\cE X}2\le P_*^{\,\bar\theta}(2P_*)^{\theta}=2^{-\bar\theta}\cdot2P_*,
\]
so the 2-norm part of $V$ shrinks by the factor $2^{-\bar\theta}$.

\emph{The 1-norm part shrinks.} The operator $Y=\cE X$ is traceless, with $x=\nrm Y1\le1$ and, by Lemma~\ref{lem:norms}, $\nrm Y2\le2^{b/2}\cdot2P_*$. If $Y\ne0$, its concentration is therefore at most $2^{b/2}\cdot2P_*/x$, and Lemma~\ref{lem:forget} gives
\[
\nrm{\Nq Y}1\le x-\delta x^{a+1};
\]
for $Y=0$, this holds trivially. The right-hand side increases with $x$ on $[0,1]$, because $(a+1)\delta\le1$, so $\nrm{\Nq Y}1\le1-\delta$.

Both parts are at most $1-\delta$, because $2^{-\bar\theta}\le1-\delta$.
\end{proof}

The two small facts used at the end, $(a+1)\delta\le1$ and $\delta\le1-2^{-\bar\theta}$, follow from one elementary inequality. With $s=1/(1-q)\ge1$, we have
\[
\delta\le c_q4^{-a}=2^{4-5s},\qquad a+1\le2s,\qquad1-2^{-\bar\theta}\ge\frac{\bar\theta}2\ge\frac{1-q}2=\frac1{2s}.
\]
Both facts thus follow from $2s\cdot2^{4-5s}\le1$, that is, $\log_2s\le5(s-1)$, which holds for all $s\ge1$. The proposition holds at every noise strength and for every channel with a budget of $b$ bits, including input-dependent unitaries.

\subsection{How large is the forgetting rate?}
\label{sm:rate}

\begin{corollary}[Rate]\label{cor:rate}
Inserting $c_q$, $a$, and $P_*$ gives
\begin{equation}
\delta^{-1}=2^{(1+4q)/(1-q)}\cdot2^{\beta_qb},\qquad\beta_q=\frac{q(1+q^2)}{(1-q)(1-q^2)} .
\label{eq:S-rate}
\end{equation}
In particular, $\delta^{-1}\le2^{b/(1-q)^2}$ for $b\ge4$, and $\delta^{-1}\le2^{(b+1)/(1-q)^2}$ for $b\ge1$.
\end{corollary}

\begin{proof}
The four factors of $\delta^{-1}=c_q^{-1}\,4^{a}\,2^{ab/2}\,P_*^{\,a}$ contribute
\[
\frac1{1-q},\qquad\frac{4q}{1-q},\qquad\frac{qb}{1-q},\qquad\frac{2q^3b}{(1-q)(1-q^2)}
\]
to $\log_2\delta^{-1}$; the first two add up to $(1+4q)/(1-q)$, the last two to $\beta_qb$. For the simple forms, compare exponents after multiplying by $(1-q)^2$, using $(1-q)^2\beta_q=q(1+q^2)/(1+q)$. The two forms become
\[
b\,(1+q+q^2)\ge(1+4q)(1+q)\qquad\text{and}\qquad b\,(1-q^3)+q(4q-3)(1+q)\ge0 .
\]
The first holds for $b\ge4$, because $4(1+q+q^2)-(1+4q)(1+q)=3-q>0$; the second holds for $b\ge1$, because $1-3q+q^2+3q^3>0$ for $0\le q\le1$.
\end{proof}

The inverse forgetting rate $\delta^{-1}$ grows by the factor $2^{\beta_q}$ for every bit of budget. At $q=0.9$, the constants are $\theta\approx0.895$, $a=18$, $c_q=2^{-10}$, and $\beta_q\approx85.7$, so $\delta^{-1}\approx2^{85.7b+46}$, compared with $2^{100b}$ from the first simple form. For $b=3$, the first simple form fails for $q>\sqrt3-1\approx0.73$, which is why the simple form of Theorem~\ref{thm:main}(i) needs $k\ge4$.

\subsection{From a shrinking distance to the memory bound}
\label{sm:memory}

\paragraph*{Washout of the purity.} An arbitrary initial state has $\log_2P_0\le n/2$. Iterating Eq.~\eqref{eq:S-purityrec} gives
\begin{equation}
\log_2P_t\le\log_2P_*+\theta^t\,\frac n2 .
\label{eq:S-washout}
\end{equation}
The excess purity of the initial state thus decays geometrically toward the cap. After $t_0=\lceil\ln n/(-\ln\theta)\rceil$ time steps, $\theta^tn\le1$, so the initial state contributes at most one pure qubit's worth, $P_t\le\sqrt2P_*$, and one more time step keeps it there, because
\[
2^{b\theta/2}\bigl(\sqrt2P_*\bigr)^\theta=2^{\theta/2}P_*\le\sqrt2P_* .
\] At $q=0.9$, $t_0=42$, $63$, and $84$ for $n=10^2$, $10^3$, and $10^4$ [dots in Fig.~\ref{figS:purity}].

\begin{proposition}[Memory bound]\label{prop:echo}
Let $\delta$ be the rate of Proposition~\ref{prop:contraction}.
\begin{enumerate}
\item[(a)] Two reservoirs started in arbitrary states $\rho$ and $\sigma$ and driven by the same inputs obey, for $L\ge t_0$,
\[
\Norm{\Phi_{u_L}\cdots\Phi_{u_1}(\rho-\sigma)}_1\le2(1-\delta)^{L-t_0}.
\]
\item[(b)] Every input history $h$ that extends into the infinite past determines a unique echo state $\rho(h)$, whose purity is at most $P_*$.
\item[(c)] If two input histories $h$ and $h'$ agree on their last $L$ inputs, then
\[
\tfrac12\Norm{\rho(h)-\rho(h')}_1\le(1-\delta)^L\le e^{-\delta L}.
\]
\end{enumerate}
\end{proposition}

\begin{proof}
(a) The difference $X_t=d(\rho_t-\sigma_t)$ is traceless and evolves by the same maps as the states, $X_t=\Nq\cE_{u_t}X_{t-1}$. After $t_0$ time steps, both states have purity at most $\sqrt2P_*$, so $\nrm{X_{t_0}}1\le2$ and $\nrm{X_{t_0}}2\le2\sqrt2P_*$, that is, $V(X_{t_0})\le2$. Proposition~\ref{prop:contraction} then gives $\nrm{X_L}1\le V(X_L)\le2(1-\delta)^{L-t_0}$.
(b) Start the reservoir at earlier and earlier times in arbitrary states. By (a), the states it reaches at a given time converge, and the limit does not depend on the initialization; this limit is the echo state. By Eq.~\eqref{eq:S-washout}, its purity is at most $P_*$.
(c) Both echo states arise from earlier echo states, which belong to the histories before the last $L$ inputs, through the same $L$ time steps. The earlier echo states have purity at most $P_*$, so their difference has $V\le2$ without any washout, and Proposition~\ref{prop:contraction} applied $L$ times gives the claim.
\end{proof}

\paragraph*{Proof of Theorem~\ref{thm:main}(i).} By Proposition~\ref{prop:echo}(c) with $b=k$, two echo states whose input histories agree on their last $L$ inputs are within trace distance $\varepsilon$ as soon as $e^{-\delta L}\le\varepsilon$. Hence $W(\varepsilon)\le\lceil\delta^{-1}\ln(1/\varepsilon)\rceil$, which with the rate of Corollary~\ref{cor:rate} is Eq.~\eqref{eq:upper}; the corollary also gives the simple form $\delta^{-1}\le2^{k/(1-q)^2}$ for $k\ge4$ and $\delta^{-1}\le2^{(k+1)/(1-q)^2}$ for $k\ge1$. The argument uses the reservoir unitary, the input encoding, and the number of qubits only through the budget, so the bound holds for every reservoir. \hfill$\square$

\paragraph*{What the memory depth bounds.} For two input histories that agree on their last $L$ inputs, a linear readout $\tr(F\rho)$ differs by at most
\[
\Norm F\,\Norm{\rho(h)-\rho(h')}_1\le2\Norm F e^{-\delta L},
\]
and the best measurement tells the two histories apart with success probability at most $\frac12(1+e^{-\delta L})$. Inputs older than $W(\varepsilon)$ thus change a readout of norm $\Norm F$ by at most $2\varepsilon\Norm F$. Large readout weights extend the reach only logarithmically: a readout with $\Norm F\le B$ sees inputs at most $\lceil\delta^{-1}\ln(2B/\varepsilon)\rceil$ time steps back at accuracy $\varepsilon$, while resolving a change of size $\varepsilon$ in $\tr(F\rho)$ from single-shot outcomes, whose variance is up to $B^2$, takes of order $(B/\varepsilon)^2$ repetitions.

Proposition~\ref{prop:echo} has further consequences. By part (a), a reservoir started in an arbitrary state is within trace distance $\varepsilon$ of the echo state at all times $t\ge t_0+\lceil\delta^{-1}\ln(1/\varepsilon)\rceil$. And a reservoir with purely unitary input encoding has a budget of zero, so every time step maps the completely mixed state to itself; by part (b), the completely mixed state is then the echo state for every input history, consistent with the observation that contractive reservoirs with input-independent fixed points carry no information about their inputs~\cite{MartinezPena2023}.

\paragraph*{Memory capacity.} The memory capacity~\cite{Jaeger2002} of a readout $y_t$ is $\mathrm{MC}=\sum_{\ell\ge1}m(\ell)$, where $m(\ell)\le1$ is the squared correlation coefficient between $y_t$ and the input $u_{t-\ell}$, maximized over the readout weights, for independent and identically distributed inputs with $0<\mathrm{Var}(u)<\infty$. A correlation coefficient does not register how weak a dependence is: a noiseless readout of a reservoir that passes its input along a chain of qubits has $m(\ell)=1$ for every $\ell$ up to the length of the chain, although the signal has decayed by $q^\ell$. A finite readout precision removes this artifact. Let $y_t=\tr(F\rho_t)+\xi_t$ with $\Norm F\le B$, where the noise $\xi_t$ (from finite measurement statistics, for instance) has, given the inputs, zero mean and variance at least $\sigma_\xi^2$, with $0<\sigma_\xi\le2B$ and finite unconditional variance, so $0<\mathrm{Var}(y_t)<\infty$. Two histories that differ only in $u_{t-\ell}$ agree on their last $\ell$ inputs, so changing $u_{t-\ell}$ alone changes $\tr(F\rho_t)$ by at most $2Be^{-\delta\ell}$. The average of $\tr(F\rho_t)$ over $u_{t-\ell}$ depends only on the other inputs and is therefore uncorrelated with $u_{t-\ell}$, while $\tr(F\rho_t)$ deviates from it by at most $2Be^{-\delta\ell}$. Hence, with $\mathrm{Var}(y_t)\ge\sigma_\xi^2$,
\[
\mathrm{Cov}(y_t,u_{t-\ell})^2\le4B^2e^{-2\delta\ell}\,\mathrm{Var}(u),\qquad m(\ell)\le\frac{4B^2e^{-2\delta\ell}}{\sigma_\xi^2}.
\]
Using $m(\ell)\le1$ for $\ell\le L=\lceil\delta^{-1}\ln(2B/\sigma_\xi)\rceil$ and summing the geometric tail beyond,
\begin{equation}
\mathrm{MC}\le L+\frac1{e^{2\delta}-1}\le\delta^{-1}\Bigl(\ln\frac{2B}{\sigma_\xi}+\frac12\Bigr)+1 .
\label{eq:S-mc}
\end{equation}
By Corollary~\ref{cor:rate}, the memory capacity of every reservoir with $k\ge4$ input qubits is thus at most $2^{k/(1-q)^2}[\ln(2B/\sigma_\xi)+\frac12]+1$: exponential in $k$, independent of $n$, and only logarithmic in the readout precision $B/\sigma_\xi$.

\subsection{How tight are the bounds?}
\label{sm:tight}

\begin{figure}[t]
\centering
\includegraphics{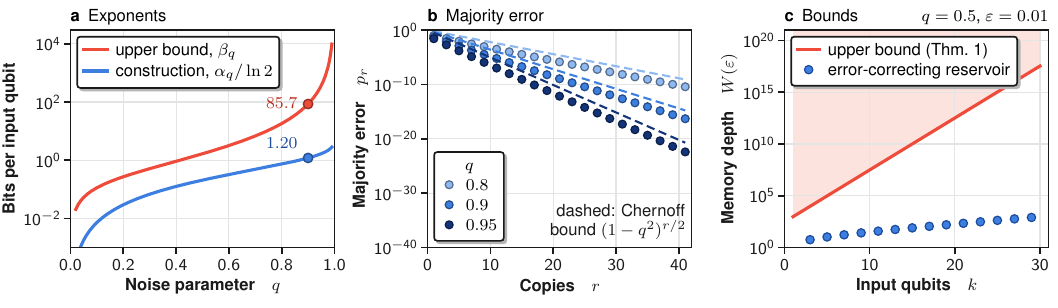}
\caption{\textbf{How tight are the bounds?}
(a)~Exponents per input qubit, in bits: $\beta_q$ of the upper bound and $\alpha_q/\ln2$ of the error-correcting reservoir; dots mark $q=0.9$.
(b)~Exact majority error $p_r$ (dots) and the Chernoff bound $(1-q^2)^{r/2}$ (dashed): the bound has the exact exponent.
(c)~Upper bound of Theorem~\ref{thm:main} with the rate~\eqref{eq:S-rate} (line; the shaded region is excluded for every reservoir) and the memory depth of the error-correcting reservoir (dots; single-qubit readout) at $q=0.5$. Both grow exponentially with $k$, at different rates.}
\label{figS:bounds}
\end{figure}

Figure~\ref{figS:bounds}(a) compares the two exponents. The exponent of the construction is the exact rate of the majority vote [Fig.~\ref{figS:bounds}(b)], so a sharper analysis of the construction cannot close the gap; the upper bound is conservative, for two reasons. Lemma~\ref{lem:forget} is a worst case over all differences with a given concentration, and its power $a=2q/(1-q)$ grows quickly as the noise weakens. Moreover, $V$ treats the 2-norm as if all of a reservoir's purity could be used to protect a single input. At $q=0.5$, the upper bound grows by a factor of $2^{1.67}$ per input qubit, $\delta^{-1}\approx2^{1.67k+6}$, and the construction by a factor of $e^{0.144}$; at $q=0.9$, the factors are $2^{85.7}$ and $e^{0.83}$ [Fig.~\ref{figS:bounds}(c)]. The scaling with $k$ is nevertheless the same, exponential on both sides, and neither bound depends on $n$.

The constants of the upper bound are conservative by many orders of magnitude. At $q=0.9$ and $\varepsilon=0.01$, the bound at $k=5$ exceeds $2^{470}$ time steps, while the error-correcting reservoir has a full-state memory depth of 316 (309 for its single-qubit readout). Theorem~\ref{thm:main} fixes the exponential dependence on $k$, not these finite-$k$ constants. Closing the gap between the exponents remains an open problem (main text, Conclusion).

\section{The washout grows with the reservoir size}
\label{sm:washoutlogn}

\begin{figure}[t]
\centering
\includegraphics{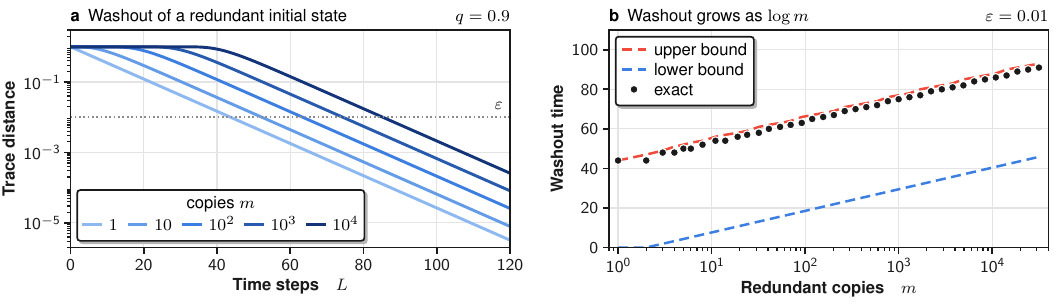}
\caption{\textbf{The washout grows with the reservoir size.} A reservoir with $U=I$ whose $m=n-k$ non-input qubits start in $\ket0^{\otimes m}$ or $\ket1^{\otimes m}$, at $q=0.9$.
(a)~Exact trace distance between the two runs: more copies keep the initial state distinguishable for longer, but every curve eventually decays at the single-qubit rate $q$.
(b)~Exact washout time to $\varepsilon=0.01$ (dots) and the two bounds derived in the text; all three grow as $\log m$.}
\label{figS:washout}
\end{figure}

A large reservoir can hold much information about its initial state, which new inputs cannot use. Take $U=I$, and let the $m=n-k$ qubits that are never reinitialized start either in $\ket0^{\otimes m}$ or in $\ket1^{\otimes m}$. After $L$ time steps, every such qubit still shows its initial value with probability $(1+q^L)/2$, independently of the others. The trace distance between the two runs therefore equals the total-variation distance between two binomial distributions, which we evaluate exactly [Fig.~\ref{figS:washout}(a)]. Simple lower and upper bounds bracket it.

\paragraph*{Lower bound.} The sum of the $Z$ outcomes has mean $\pm mq^L$ and variance at most $m$. By Chebyshev's inequality, deciding by majority errs with probability at most $1/(mq^{2L})$ under either initialization, so the trace distance is at least $1-2/(mq^{2L})$. It stays above $\varepsilon$ as long as $mq^{2L}>2/(1-\varepsilon)$, that is, for
\[
L<\frac{\ln[m(1-\varepsilon)/2]}{2|\ln q|}.
\]
A washout that grows as $\log m$ is therefore unavoidable.

\paragraph*{Upper bound.} The two product states have fidelity $(1-q^{2L})^{m/2}$, so their trace distance is at most $\sqrt m\,q^L$. It falls below $\varepsilon$ after
\[
\Bigl\lceil\frac{\ln(\sqrt m/\varepsilon)}{|\ln q|}\Bigr\rceil
\]
time steps: copies that are never repaired become indistinguishable after $O_q(\log m)$ time steps.

Both bounds and the exact washout time grow as $\log m$ [Fig.~\ref{figS:washout}(b)]. The same estimate shows that $r$ unrepaired copies of an input keep it readable for only $O_q(\log r)$ time steps. The memory of new inputs, in contrast, is bounded independently of $n$.

\section{The error-correcting reservoir}
\label{sm:rep}

This section derives the exact response of the reservoir of the main text and the lower bound of Theorem~\ref{thm:main}, including the origin of the exponent $\alpha_q$.

\subsection{The construction}

The reservoir has $n=2r+2$ qubits: an input register of $k=r+2$ qubits, which consists of two input qubits $(c,b)$ and $r$ ancillas, and $r$ memory qubits. A blank input, which asks the reservoir to keep its memory, injects $\ket{c,b}=\ket{0,0}$; a store input for the bit $b$ injects $\ket{1,b}$; the ancillas are always reinitialized in $\ket0^{\otimes r}$. The reservoir unitary is the permutation of computational basis states
\begin{equation}
U\ket{z,c,b,x}=\ket{x,\,c,\,b,\,z\oplus F_{c,b}(x)\bm1},
\label{eq:S-repU}
\end{equation}
where $z$ and $x$ are the contents of the ancillas and of the memory qubits, $\bm1=(1,\dots,1)$, $F_{0,b}(x)=\mathrm{maj}(x)$, and $F_{1,b}(x)=b$. That is, $U$ writes either the majority of the stored copies (blank input) or the new bit (store input) into the fresh ancillas, and it swaps ancillas and memory qubits. The map is invertible, because its output reveals $x$, $c$, and $b$, hence $F_{c,b}(x)$, and hence $z$; a permutation of basis states is a unitary. The construction is classical reversible logic run on qubits, and it uses no measurements and no feedback: majority-vote error correction of a repetition code~\cite{vonNeumann1956}, carried out unitarily with fresh ancillas. This is measurement-free error correction~\cite{Heussen2024} in its simplest form.

\subsection{Exact response}

Under depolarizing noise, a qubit in $\ket0$ or $\ket1$ flips with probability $p=(1-q)/2$ and is otherwise left alone. Diagonal states thus stay diagonal, and every qubit behaves as a classical bit. Let $\hat b_j$ denote the bit that the unitary writes into the ancillas at time step $j$. During a blank time step, $\hat b_j=\mathrm{maj}(x)$, where $x$ are the copies written at time step $j-1$, each of which has flipped with probability $p$ during one round of noise, independently of the others. Hence $\Pr[\hat b_j\ne\hat b_{j-1}]=p_r$ for either value of $\hat b_{j-1}$, and
\begin{equation}
\mathbb E\bigl[(-1)^{\hat b_j}\bigr]=\lambda_r\,\mathbb E\bigl[(-1)^{\hat b_{j-1}}\bigr],\qquad\lambda_r=1-2p_r .
\end{equation}
A stored bit sets $\hat b=b$ exactly, after which the bias decays as $\lambda_r^{\,t}$. Each memory qubit equals the current bit $\hat b$ up to one round of noise, which contributes the factor $q$ in Eq.~\eqref{eq:response}.

\subsection{The majority error and the exponent \texorpdfstring{$\alpha_q$}{alpha q}}

The majority of $r$ copies, for odd $r$, is wrong if more than half of them have flipped. Write $\xi_i=+1$ if copy $i$ is intact and $\xi_i=-1$ if it has flipped, so that the $\xi_i$ are independent with mean $q$. For every $\eta>0$, Markov's inequality applied to $e^{-\eta\sum_i\xi_i}$ gives the Chernoff bound
\begin{equation}
p_r=\Pr\Bigl[\sum_i\xi_i<0\Bigr]\le\Bigl[\tfrac{1+q}2e^{-\eta}+\tfrac{1-q}2e^{\eta}\Bigr]^r .
\end{equation}
The choice $e^{2\eta}=(1+q)/(1-q)$ minimizes the bracket, which then equals $\sqrt{1-q^2}$. Hence
\begin{equation}
p_r\le(1-q^2)^{r/2}=e^{-\alpha_qr},\qquad\alpha_q=\tfrac12\ln\frac1{1-q^2},
\label{eq:S-chernoff}
\end{equation}
the bound quoted in the main text. The exponent $\alpha_q$ is the rate at which the majority vote becomes reliable as copies are added. It is exact, since $p_{r+2}/p_r\to1-q^2$ for large $r$; the Chernoff bound overestimates $p_r$ only by a factor polynomial in $r$ [Fig.~\ref{figS:bounds}(b)].

\begin{figure}[t]
\centering
\includegraphics{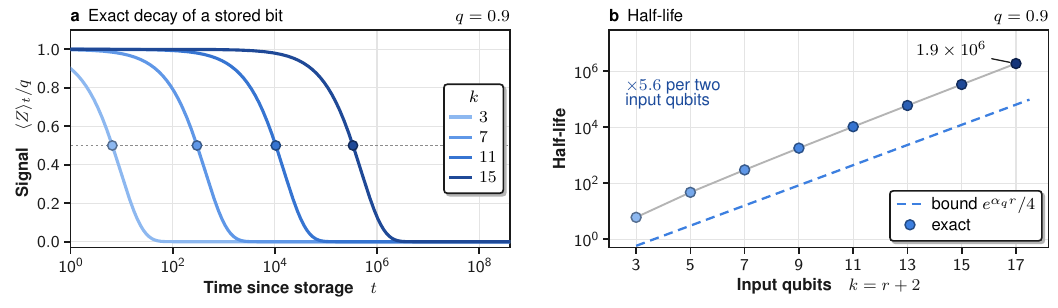}
\caption{\textbf{Exact response of the error-correcting reservoir} at $q=0.9$.
(a)~Decay of a stored bit, $\langle Z\rangle_t/q=\lambda_r^{\,t}$; dots mark the half-lives.
(b)~Exact half-life versus the number of input qubits (dots) and the bound $e^{\alpha_qr}/4$ (dashed). Every two input qubits multiply the half-life by a nearly constant factor.}
\label{figS:rep}
\end{figure}

By Bernoulli's inequality,
\[
\lambda_r^{\,t}\ge1-2tp_r\ge\tfrac12\qquad\text{for }t\le e^{\alpha_qr}/4,
\]
so the half-life of a stored bit is at least $\lfloor e^{\alpha_qr}/4\rfloor$. Figure~\ref{figS:rep} shows the exact decay and half-lives at $q=0.9$, and Table~\ref{tabS:halflife} lists exact half-lives for several noise strengths. From $k=3$ to $k=17$, the half-lives at $q=0.9$ are $6$, $47$, $298$, $1790$, $10\,431$, $59\,739$, $337\,940$, and $1\,894\,286$ time steps: seventeen input qubits per time step keep a stored bit readable for almost two million time steps, and every two additional input qubits multiply this time by more than five.

\begin{table}[t]
\caption{Exact half-lives, in time steps, of a bit stored in the error-correcting reservoir: the largest $T$ with $\lambda_r^T\ge1/2$, for $k=r+2$ input qubits.}
\label{tabS:halflife}
\begin{ruledtabular}
\begin{tabular}{cccccc}
$q$ & $k=3$ & $k=5$ & $k=9$ & $k=13$ & $k=17$ \\ \hline
0.8 & 3 & 12 & 126 & 1171 & 10\,306 \\
0.9 & 6 & 47 & 1790 & 59\,739 & $1.9\times10^{6}$ \\
0.95 & 13 & 187 & 26\,931 & $3.4\times10^{6}$ & $4.1\times10^{8}$ \\
0.99 & 68 & 4636 & $1.6\times10^{7}$ & $4.9\times10^{10}$ & $1.4\times10^{14}$ \\
\end{tabular}
\end{ruledtabular}
\end{table}

\subsection{Proof of Theorem~1(ii)}
\label{sm:lowerbound}

Two input histories that differ only in the bit stored $L$ time steps ago, followed by blanks, agree on their last $L$ inputs. Because a stored bit overwrites the memory qubits regardless of their previous content, both histories lead to valid echo states, and the memory is not a remnant of the initialization. By Eq.~\eqref{eq:response}, the two echo states give expectation values of $Z$ on a memory qubit that differ by $2q\lambda_r^L$. Since $|\tr[O(\rho-\rho')]|\le\Norm O\,\Norm{\rho-\rho'}_1$, their trace distance is at least $q\lambda_r^L$, which exceeds $\varepsilon$ as long as $L<\ln(q/\varepsilon)/(-\ln\lambda_r)$. Hence
\begin{equation}
W(\varepsilon)\ge\frac{\ln(q/\varepsilon)}{-\ln\lambda_r}\ge\frac14\,e^{\alpha_qr}\ln\frac q\varepsilon .
\label{eq:S-lower}
\end{equation}
The factor $q$ comes from the readout qubit's own last round of noise. The second inequality needs $\varepsilon<q$, so that $\ln(q/\varepsilon)>0$, and the inequality $-\ln(1-x)\le2x$ for $0\le x\le\frac12$, applied to $x=2p_r$. This requires $p_r\le\frac14$, and then Eq.~\eqref{eq:S-chernoff} gives
\[
-\ln\lambda_r\le4p_r\le4e^{-\alpha_qr}.
\]

For a general number $k$ of input qubits, choose $r$ as the largest odd number with $r\le k-2$, and leave the remaining input qubits unused. Then $r\ge k-3$ and $n=k+r\le2k$. If $e^{\alpha_q(k-3)}\ge4$, then $p_r\le e^{-\alpha_qr}\le\frac14$, and Eq.~\eqref{eq:S-lower} gives Eq.~\eqref{eq:lower}. At $q=0.9$, the condition holds for $k\ge5$. \hfill$\square$

\subsection{Further properties}

\paragraph*{The single-qubit readout is nearly optimal.} The lower bound uses the readout of a single memory qubit. The full reservoir state carries slightly more information: for $k=5$ and $q=0.9$, the exact trace distance between the two echo states [Fig.~\ref{figS:sim}(a)] gives a memory depth of 316 time steps at $\varepsilon=0.01$, compared with 309 for the single-qubit readout of Eq.~\eqref{eq:S-lower}.

\paragraph*{Without repair.} The same reservoir without the majority vote writes the bit into its $r$ memory qubits and then leaves them alone. After $t$ further time steps, every copy has the Bloch vector $\pm q^{t+1}\hat z$, and the trace distance between the two echo states is the binomial sum
\[
\frac12\sum_{j=0}^{r}\binom rj\bigl|a^jb^{r-j}-b^ja^{r-j}\bigr|,\qquad a,b=\frac{1\pm q^{t+1}}2,
\]
which no readout can beat. At $q=0.9$ and $\varepsilon=0.01$, the memory depth of the copies grows only from 43 time steps for $r=1$ to 53 for $r=13$, whereas the error-correcting reservoir with $r=13$ copies ($k=15$) remembers $2.2\times10^6$ time steps [Fig.~\ref{fig:attain}(b)]. The fresh qubits extend the memory only because the reservoir uses them for repair.

\paragraph*{Retention cost.} Suppose that a reservoir must keep the expectation values of an observable $O$ with $\Norm O\le1$ for a stored 0 and a stored 1 at least $q$ apart for $T$ time steps. Proposition~\ref{prop:echo}(c) bounds this gap by $2e^{-\delta T}$, so $T\le\delta^{-1}\ln(2/q)$, and with Eq.~\eqref{eq:S-rate}, $k=\Omega_q(\log T)$. Conversely, Eq.~\eqref{eq:response} keeps the gap for all $t\le T$ if $T\le e^{\alpha_qr}/4$. An odd $r\ge\ln(4T)/\alpha_q$ and $k=r+2$ therefore suffice, so $k=O_q(\log T)$. Thus $\Theta_q(\log T)$ reinitialized qubits per time step are necessary and sufficient to retain an input for $T$ time steps.

\paragraph*{Anisotropic noise.} For independent unital noise $\mathcal M$ on every qubit, the bit is best stored along the Bloch axis that $\mathcal M$ preserves best. Let $s_{\mathcal M}$ be the largest singular value of the matrix by which $\mathcal M$ acts on Bloch vectors, with unit vectors $\bm v$ and $\bm v'$ such that $\mathcal M$ maps the Bloch vector $\bm v$ to $s_{\mathcal M}\bm v'$. Conjugating the memory block of Eq.~\eqref{eq:S-repU} by single-qubit rotations from the $\bm v'$ basis and into the $\bm v$ basis makes every memory qubit an independent copy of the stored bit, flipped with probability $(1-s_{\mathcal M})/2$. The lower bound then holds with $q$ replaced by $s_{\mathcal M}$.

\paragraph*{Non-Clifford dynamics.} The majority vote is a non-Clifford operation, as long-term memory requires~\cite{Nelson2026} (Sec.~\ref{sm:prior}).

\paragraph*{Memory versus computation.} A long memory does not imply computational power or classical hardness: a single number, the bias $\lambda_r^{\,t}$, describes the entire response of this reservoir.

\section{Entropy budget beyond depolarizing noise}
\label{sm:budget}

This section extends Theorem~\ref{thm:main} to more general noise, using the budget of Sec.~\ref{sm:norms}, and shows where the extension ends. Table~\ref{tabS:budget} summarizes the rule.

\begin{table}[t]
\caption{Entropy budget of common ingredients of quantum reservoirs: how many bits of purity a process can inject per time step. The memory depth is at most exponential in the total budget, and processes with zero budget cannot raise the memory limit. Processes outside the model are not covered by the bound.}
\label{tabS:budget}
\begin{ruledtabular}
\begin{tabular}{>{\raggedright\arraybackslash}p{5.3cm}>{\raggedright\arraybackslash}p{2.7cm}>{\raggedright\arraybackslash}p{8.3cm}}
Process & Budget per time step & Consequence for the memory \\ \hline
Reinitialization of $k$ input qubits & $k$ bits & memory depth $e^{\Theta(k)}$ under depolarizing noise (Theorem~\ref{thm:main}) \\
Measurement followed by reset of $m$ qubits & $m$ bits & counts like $m$ reinitialized input qubits~\cite{Yasuda2023,Hu2024} \\
Unitary input encoding (input-dependent gates) & 0 & without another entropy sink, the echo state is the completely mixed state \\
Projective (or other L\"uders) measurement of reservoir qubits whose outcomes only enter the readout & 0 & the averaged measurement channel is unital and cannot raise the memory limit~\cite{Mujal2023}; general instruments count with their own budget, and indirect measurements through fresh meter qubits count like reinitialized qubits~\cite{MorguiSancho2026} \\
Unital noise (Pauli errors, dephasing, coherent errors, crosstalk) & 0 & does not raise the memory upper bound \\
Amplitude damping with probability $0<\gamma<1$ on all $n$ qubits & $n\log_2(1+\gamma)$ bits & memory depth can grow exponentially with $n$, even for $k=2$~\cite{BenOr2013,Monzani2024,Freiheit2026} \\
Feedback of measurement outcomes; non-Markovian noise; noise without a depolarizing component & outside the model & memory can be unbounded (Sec.~\ref{sm:counter})~\cite{Sannia2026} \\
\end{tabular}
\end{ruledtabular}
\end{table}

\subsection{Unital noise and measurements}

Let the noise at time step $t$ be $\mathcal R_t\circ\Nq\circ\mathcal S_t$, where $\mathcal R_t$ and $\mathcal S_t$ are unital channels. They can be correlated across qubits, coherent, input dependent, or drifting in time. This case reduces to the one already proven. First, $\mathcal S_t$ can be absorbed into the injection step: $\mathcal S_t\circ\cE_u$ is a channel, and because $\mathcal S_t$ is positive and unital,
\[
\mathcal S_t\cE_u(I)\le2^k\mathcal S_t(I)=2^kI,
\]
so it has the same budget of $k$ bits. Second, $\mathcal R_t$ cannot increase $V$, because a unital channel contracts both norms (Lemma~\ref{lem:norms} with $b=0$). Hence
\[
V(\mathcal R_t\Nq\mathcal S_t\cE_uX)\le V(\Nq\mathcal S_t\cE_uX)\le(1-\delta)\,V(X),
\]
and the same two facts give the purity recursion. Two input histories that agree on their last $L$ inputs share every map during those $L$ time steps, so Theorem~\ref{thm:main} holds verbatim; its lower bound needs only $\mathcal R_t=\mathcal S_t=\mathrm{id}$.

Measurements are covered in the same way. A measurement whose post-measurement states follow the L\"uders rule, with Kraus operators $\sqrt{E_i}$, is unital when its outcomes are discarded, because $\sum_i\sqrt{E_i}\sqrt{E_i}=\sum_iE_i=I$; this includes every projective measurement. Such measurements of the reservoir qubits whose outcomes only enter the readout therefore cannot raise the memory limit. A general measurement instrument with Kraus operators $A_i$, $\sum_iA_i^\dagger A_i=I$, is unital only if also $\sum_iA_iA_i^\dagger=I$; the Kraus operators of amplitude damping are a counterexample. A general instrument therefore counts with its budget $b=\log_2\Norm{\sum_iA_iA_i^\dagger}$, like any other channel. An indirect measurement through a meter qubit that is freshly prepared at every time step is of this kind: the meter is a reinitialized qubit and adds one bit of budget per time step, consistent with the role of measurement back-action as a source of dissipation~\cite{MorguiSancho2026}.

\subsection{Non-unital noise and resets}

Now let $\mathcal R_t$ and $\mathcal S_t$ be arbitrary channels with budgets of $b_R$ and $b_S$ bits, with $\mathcal R_t$ independent of the inputs. Regroup the time steps as $\rho_t=\mathcal R_t(\tilde\rho_t)$, with
\begin{equation}
\tilde\rho_t=\Psi_t(\tilde\rho_{t-1}),\qquad\Psi_t=\Nq\circ\mathcal S_t\circ\cE_{u_t}\circ\mathcal R_{t-1},
\end{equation}
and $\mathcal R_0=\mathrm{id}$. By positivity, $\mathcal S_t\cE_{u_t}\mathcal R_{t-1}(I)\le2^{k+b_R+b_S}I$, so Sec.~\ref{sm:upper} applies to $\tilde\rho_t$ with the budget $b=k+b_R+b_S$. Two input histories that agree on their last $L$ inputs share every $\Psi_t$ during those $L$ time steps. The final channel $\mathcal R_t$ cannot increase trace distances, so the memory depth of $\rho_t$ is at most that of $\tilde\rho_t$. Apart from the washout time, the reservoir size thus enters only through $b_R+b_S$. The bounds hold for every realization of random channels $\mathcal R_t$ and $\mathcal S_t$, and hence also on average.

The main text uses the budgets of resets and of amplitude damping. A reset channel $\rho\mapsto\tr(\rho)\ketbra00$ maps $I$ to $2\ketbra00\le2I$, so a reset of $m$ qubits has a budget of $m$ bits. Amplitude damping $\AD_\gamma$ with probability $\gamma$ maps $I$ to $\mathrm{diag}(1+\gamma,1-\gamma)\le(1+\gamma)I$, so a layer of amplitude damping on all $n$ qubits has a budget of $n\log_2(1+\gamma)$ bits.

\subsection{Every unital qubit channel with full-rank Choi matrix contains depolarizing noise}

Let $\mathcal M$ be a unital qubit channel, and let
\[
\Omega_{\mathcal M}=\sum_{i,j=0}^1\mathcal M(\ketbra ij)\otimes\ketbra ij
\]
be its Choi matrix, a positive operator with $\tr\Omega_{\mathcal M}=2$. Complete depolarization, $\cD_0(X)=\tr(X)I/2$, has $\Omega_{\cD_0}=I/2$. Suppose that $\Omega_{\mathcal M}$ is positive definite and that $\mathcal M\ne\cD_0$. Then the smallest eigenvalue of $\Omega_{\mathcal M}$ lies in $(0,1/2)$, and $q_{\mathcal M}=1-2\lambda_{\min}(\Omega_{\mathcal M})\in(0,1)$. The map
\[
\mathcal T=\frac{\mathcal M-(1-q_{\mathcal M})\,\cD_0}{q_{\mathcal M}}
\]
is trace preserving and unital, and its Choi matrix $[\Omega_{\mathcal M}-\lambda_{\min}(\Omega_{\mathcal M})I]/q_{\mathcal M}$ is positive semidefinite, so $\mathcal T$ is a channel. Because $\mathcal T\circ\cD_q=q\mathcal T+(1-q)\cD_0$ for every unital channel $\mathcal T$, we obtain
\begin{equation}
\mathcal M=\mathcal T\circ\cD_{q_{\mathcal M}} .
\end{equation}
Every such channel is thus depolarizing noise of strength $q_{\mathcal M}$ followed by a unital channel, which is the form covered above. For a Pauli channel with probabilities $p_I,p_X,p_Y,p_Z$, the Choi eigenvalues are $2p_I,\dots,2p_Z$, so $1-q_{\mathcal M}=4\min\{p_I,p_X,p_Y,p_Z\}$. For example, Pauli noise with $p_X=p_Y=0.01$ and $p_Z=0.08$ has $q_{\mathcal M}=0.96$.

\begin{figure}[t]
\centering
\includegraphics{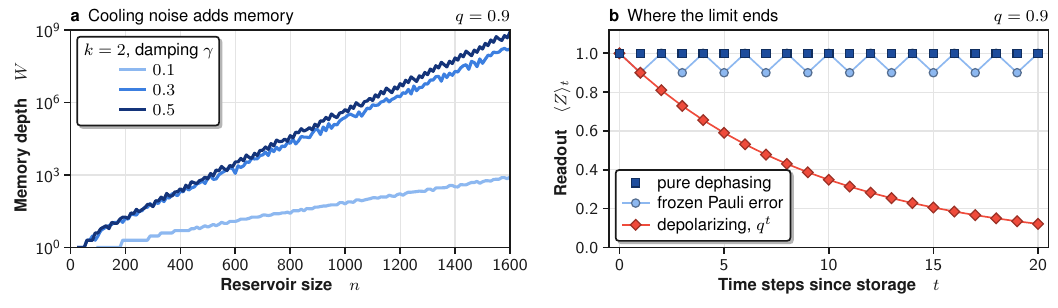}
\caption{\textbf{The entropy budget at work.}
(a)~Cooling noise adds memory. Lower bound $\lfloor1/(4p_{\mathrm f})\rfloor$ on the memory depth of the cooling reservoir of Sec.~\ref{sm:cooling}, with only $k=2$ input qubits, under the noise $(\cD_q\circ\AD_\gamma)^{\otimes n}$; $p_{\mathrm f}$ is the worst-case probability of losing the stored bit per time step, from exact binomial tails.
(b)~Where the limit ends. Readout of a stored bit under pure dephasing (no decay), under a Pauli error that is drawn once and then repeated at every time step (perfect revivals, although one time step is depolarizing on average), and under Markovian depolarizing noise ($q^t$).}
\label{figS:budget}
\end{figure}

\subsection{Cooling noise: memory exponential in \texorpdfstring{$n$}{n}}
\label{sm:cooling}

Amplitude damping can make the memory grow with the reservoir size, even with two input qubits. The construction uses idle qubits as a refrigerator~\cite{BenOr2013}. Let every time step end with the noise $(\cD_q\circ\AD_\gamma)^{\otimes n}$, with $0<\gamma<1$.

\paragraph*{Idle qubits cool.} On diagonal states, one time step of noise acts on every qubit as a bit channel: a qubit that is 1 with probability $x$ becomes 1 with probability
\[
p+\kappa x,\qquad p=\frac{1-q}2,\qquad\kappa=q(1-\gamma).
\] Here $0<\kappa<1$ is the fraction of a qubit's polarization that survives one time step. The condition $\gamma<1$ is necessary: for $\gamma=1$, the noise resets every qubit to $\ket0$ before depolarizing it, the state after every time step is $[(I+qZ)/2]^{\otimes n}$ for all inputs, and the memory depth is zero. A qubit left alone approaches the stationary probability $p_\infty=p/(1-\kappa)$ of being 1, which is below $1/2$ for every $\gamma>0$: idle qubits cool. After $s$ time steps, starting from either value, a qubit is 1 with probability at most
\[
\mu_s=p_\infty+\kappa^s(1-p_\infty),
\]
and we choose the cooling time $s$ such that $\mu_s<1/2$.

\paragraph*{The reservoir.} The reservoir has the two input qubits $(c,b)$ of Sec.~\ref{sm:rep} and $s+1$ blocks $x_0,\dots,x_s$ of $r=\lfloor(n-2)/(s+1)\rfloor$ qubits each. Block $x_0$ holds the stored bit, and the other blocks form a queue in which qubits cool. The reservoir unitary is the basis permutation
\begin{equation}
U\ket{c,b}\ket{x_0,x_1,\dots,x_s}=\ket{c,b}\ket{x_s\oplus F_{c,b}(x_0)\bm1,\,x_0,\dots,x_{s-1}} .
\end{equation}
For a store input, $F_{1,b}(x)=b$. For a blank input, $F_{0,b}(x)=1$ if the fraction of ones in $x$ is at least the threshold $\chi=p+\kappa/2$, and $F_{0,b}(x)=0$ otherwise. In effect, $U$ decodes the stored bit by a threshold vote, writes it into the coolest block $x_s$, and moves the other blocks one position along the queue.

\paragraph*{Error per time step.} The reused block has cooled for $s$ time steps, so each of its bits differs from the newly written bit $F$ with probability at most $\mu_s$, independently given the past. After the noise, each bit of the new memory block is therefore 1 with probability at least $p+\kappa(1-\mu_s)$ if $F=1$, and at most $p+\kappa\mu_s$ if $F=0$. The threshold $\chi$ lies between these two values, at a distance of at least $\kappa(1/2-\mu_s)$ from both. The next vote therefore errs with probability at most
\begin{equation}
p_{\mathrm f}=\max\bigl\{\Pr[\mathrm{Bin}(r,p+\kappa\mu_s)\ge\chi r],\ \Pr[\mathrm{Bin}(r,p+\kappa(1-\mu_s))<\chi r]\bigr\},
\end{equation}
and Hoeffding's inequality~\cite{Hoeffding1963} gives $p_{\mathrm f}\le e^{-\nu r}$ with $\nu=2\kappa^2(1/2-\mu_s)^2$. The failure probability thus falls exponentially with the block size.

\paragraph*{Memory.} Consider two input histories that differ only in the bit stored $L$ time steps ago, followed by blanks. The stored bit survives the $L+1$ subsequent votes with probability at least $1-(L+1)p_{\mathrm f}$, so the threshold observable of block $x_0$, $+1$ if the fraction of ones is at least $\chi$ and $-1$ otherwise, differs between the two echo states by at least $2-4(L+1)p_{\mathrm f}$. For $(L+1)p_{\mathrm f}\le1/4$, this difference is at least $1$, and the trace distance is at least $1/2$. Hence
\begin{equation}
W(\varepsilon)\ge\Bigl\lfloor\frac1{4p_{\mathrm f}}\Bigr\rfloor\ge\Bigl\lfloor\frac{e^{\nu r}}4\Bigr\rfloor\qquad\text{for all }\varepsilon<\frac12 .
\end{equation}
With only two input qubits, the memory depth is exponential in $r\approx n/(s+1)$, and hence in $n$ [Fig.~\ref{figS:budget}(a)].

This is the classical counterpart of the quantum refrigerator of Ref.~\cite{BenOr2013}; for fault-tolerant quantum computation under amplitude damping, the achievable length is known to be at most exponential in $n$~\cite{Fawzi2022}.

\paragraph*{Numbers.} At $q=0.9$ and $\gamma=0.3$, we have $p=0.05$, $\kappa=0.63$, and $p_\infty\approx0.135$. The exact binomial tails, optimized over the cooling time $s$, push $p_{\mathrm f}$ below $10^{-6}$ at $n=1010$ qubits, with $s=7$ and $r=126$. The cooling route is expensive: without damping, the error-correcting reservoir reaches a failure probability of $1.8\times10^{-7}$ per time step with 32 qubits, 17 of them input qubits. On hardware dominated by energy relaxation, the reservoir size does matter for memory, through cooling, consistent with the memory gains attributed to amplitude damping in gate-based reservoirs~\cite{Monzani2024} and with collective relaxation as a scalable fading-memory mechanism~\cite{Freiheit2026}.

\subsection{Where the limit ends: two counterexamples}
\label{sm:counter}

Theorem~\ref{thm:main} assumes that the noise has a depolarizing component and that it is Markovian, and neither assumption can be dropped. Consider a single memory qubit next to the two input qubits $(c,b)$. The reservoir unitary swaps $b$ into the memory qubit when $c=1$ and acts trivially when $c=0$; the old memory value leaves with the input register at the next input injection.

\paragraph*{Pure dephasing.} Under the dephasing channel $\mathcal Z_\phi(X)=(1-\phi)X+\phi\,ZXZ$, computational-basis states are fixed points, so a stored bit stays perfectly readable forever.

\paragraph*{Frozen Pauli errors.} Let each qubit carry a Pauli operator $S\in\{I,X,Y,Z\}$, drawn once with probabilities $\frac{1+3q}4,\frac{1-q}4,\frac{1-q}4,\frac{1-q}4$ and then applied at every time step. Averaged over the draw, one time step is exactly $\cD_q^{\otimes n}$, because
\begin{equation}
\mathbb E[S\rho S]=\tfrac{1+3q}4\rho+\tfrac{1-q}4(X\rho X+Y\rho Y+Z\rho Z)=\cD_q(\rho).
\end{equation}
But $S^2=I$, so after an even number of time steps the memory qubit is back in $\ketbra bb$, and the stored bit is read perfectly at every second time step. Averaged over the draw, the readout alternates between $1$ and $q$ [Fig.~\ref{figS:budget}(b)]. Here the environment remembers the noise, which is why the theorem requires Markovian noise; non-Markovian dynamics can likewise extend the memory of quantum reservoirs in practice~\cite{Sannia2026}.

In both cases, the reservoir has no unique echo state, and no finite memory depth exists for input histories that contain a stored bit. Finally, a reservoir whose measurement outcomes are stored and fed back is coupled to a classical memory outside the model, which is an entropy sink of its own.

\section{Numerical simulations}
\label{sm:numerics}

\subsection{Reservoirs}

We simulate the density matrix of Eq.~\eqref{eq:model} exactly for $n\le9$ qubits. The first $k$ qubits form the input register. Each input qubit is reinitialized in $\ket{\psi(u_t)}=\sqrt{1-u_t}\,\ket0+\sqrt{u_t}\,\ket1$~\cite{Fujii2017}, with independent inputs $u_t$ drawn uniformly from $[0,1]$. We consider two reservoir unitaries. The \emph{Ising reservoir} evolves for a time $\Delta t$ under the transverse-field Ising Hamiltonian
\begin{equation}
H=\sum_{i<j}J_{ij}X_iX_j+g\sum_iZ_i,
\end{equation}
with couplings $J_{ij}$ drawn uniformly from $[-J/2,J/2]$, transverse field $g=J$, and $J\Delta t=10$, as in the standard model of QRC~\cite{Fujii2017,Nakajima2019,MartinezPena2021}. The \emph{Haar reservoir} uses a unitary drawn from the Haar measure on $U(2^n)$, the extreme of information scrambling. After every time step, each qubit undergoes depolarizing noise $\cD_q$. Table~\ref{tabS:sim} lists all parameters.

\begin{table}[t]
\caption{Parameters of the simulations in Fig.~\ref{fig:attain} of the main text and Figs.~\ref{figS:sim} and~\ref{figS:cold}.}
\label{tabS:sim}
\begin{ruledtabular}
\begin{tabular}{>{\raggedright\arraybackslash}p{4.2cm}>{\raggedright\arraybackslash}p{7.2cm}}
Parameter & Value \\ \hline
Noise parameter $q$ & $0.9$; also $0.95$--$1$ [Fig.~\ref{figS:sim}(b)] \\
Accuracy $\varepsilon$ & $0.01$ \\
Reservoir size $n$ & 3--9 \\
Input qubits $k$ & 1--4 \\
Ising couplings & $J_{ij}\in[-J/2,J/2]$, field $g=J$ \\
Evolution time & $J\Delta t=10$; also 0.3--30 [Fig.~\ref{figS:sim}(c)] \\
Input encoding & $\sqrt{1-u}\,\ket0+\sqrt u\,\ket1$, $u\in[0,1]$; cold test: additional input qubits in $\ket0$ \\
Washout & until three initial states agree within $10^{-6}$ (60--2360 time steps) \\
History pairs & one differing input; all earlier inputs differing \\
Continuations & 5 per history pair \\
Realizations & 6 reservoirs per $n$ and $k$, the same for all $q$, $J\Delta t$, and encodings \\
Stopping criterion & $D(L)<3\times10^{-3}$ \\
\end{tabular}
\end{ruledtabular}
\end{table}

\subsection{Memory depth}

The memory depth $W(\varepsilon)$ of the main text is a supremum over all pairs of input histories that agree on their last $L$ inputs. Because each time step is linear, the difference $\Delta_L$ of the two reservoir states evolves under the same map as the states themselves, and we follow it directly. We evaluate two kinds of history pairs for every reservoir. In the first, the histories differ in a single input: we prepare the echo state of a random input history and inject the two most different inputs, $u=0$ and $u=1$, into two copies of it. In the second, the histories differ in all earlier inputs: we prepare the echo states of the constant histories $u=0$ and $u=1$. Each pair is followed by five common continuations, two random input sequences and the constant inputs $u=0$, $1/2$, and $1$, and we record $D(L)=\frac12\Norm{\Delta_L}_1$ after $L$ common time steps until $D(L)<3\times10^{-3}$, well below $\varepsilon$. The memory depth of a reservoir is $1+\max\{L:D(L)>\varepsilon\}$, maximized over both pairs and all continuations; as a maximum over a subset of all pairs, it can only underestimate the supremum, and check (iv) below tests it against a much larger set.

Histories that differ in all earlier inputs set the memory depth of almost every reservoir: in the main runs [Figs.~\ref{fig:attain}(c) and~\ref{figS:cold}], they give a larger value than a single differing input in 94 of 180 cases, the same value in 81, and a smaller one in 5. Their differences accumulate in the slowest modes of the reservoir, which a single input excites only weakly; for the Ising reservoirs with one input qubit, they add about five time steps. The idle qubit, a single qubit that holds the input without repair, has memory depth $1+\max\{L:q^{L+1}>\varepsilon\}=43$ at $q=0.9$ and $\varepsilon=0.01$.

\subsection{Validation}

The code and the estimator pass the following checks. (i) The evolution of differences agrees with the difference of evolved states to machine precision. (ii) For the error-correcting reservoir, the simulated single-qubit readout reproduces the exact response $q\lambda_r^L$ of Eq.~\eqref{eq:response} to machine precision for $r=1$ and $r=3$. (iii) Every echo state is prepared by a washout that continues until runs from the completely mixed state and from the product states $\ket{0\cdots0}$ and $\ket{1\cdots1}$, driven by the same inputs, agree within a trace distance of $10^{-6}$, far below $\varepsilon$. This takes at most 130 time steps at $q=0.9$ and up to 2360 time steps as $q\to1$, including the noiseless limit, which the theorem does not cover; there, the three runs agree for every simulated reservoir. (iv) A much larger set of history pairs changes the memory depth by at most one time step. For 24 reservoirs (Ising and Haar, $n=6$ and $8$, $k=1$ and $2$), we compared the estimator with blocks of 2 to 30 differing inputs, all 32 sign patterns of six differing inputs up to exchanging the two histories, constant histories with the input pairs $(0,\frac12)$, $(\frac12,1)$, and $(\frac14,\frac34)$, three pairs of independent random histories, and continuations that maximize the trace distance greedily at every time step.

\begin{figure}[t]
\centering
\includegraphics{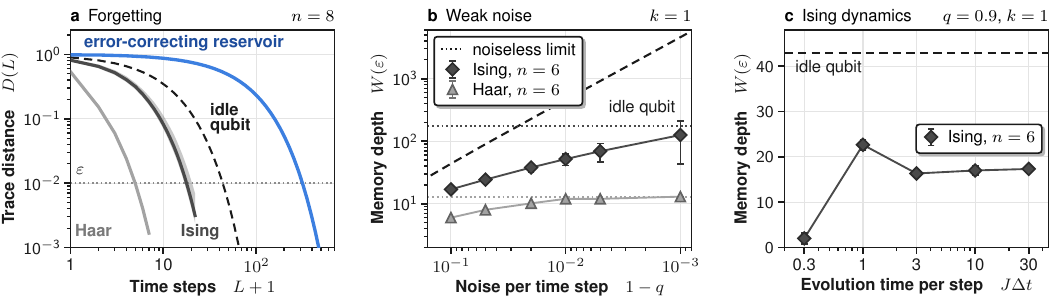}
\caption{\textbf{Simulations of standard reservoirs.}
(a)~Trace distance $D(L)$ between the echo states of two input histories that agree on their last $L$ inputs, for reservoirs with $n=8$ qubits: Haar and Ising reservoirs with one input qubit, for histories that differ in all earlier inputs (median and range over six reservoirs, first random continuation), an idle qubit (dashed), and the error-correcting reservoir with $k=5$, for histories that differ in the stored bit (blue). The memory depth is where $D(L)$ falls below $\varepsilon$ (dotted).
(b)~Memory depth of the same six reservoirs with $n=6$ and $k=1$ versus the noise per time step (mean and standard deviation), compared with an idle qubit; dotted lines mark the noiseless limit.
(c)~Memory depth of the six Ising reservoirs with $n=6$ versus the evolution time per time step, with fixed couplings.}
\label{figS:sim}
\end{figure}

\subsection{Results and robustness}
\label{sm:results}

The forgetting curves of Fig.~\ref{figS:sim}(a) show the hierarchy of the main text directly: the Haar reservoir forgets within a few time steps, the Ising reservoir within about 18, the idle qubit within 43, and the error-correcting reservoir with the same $n=8$ only after 316 time steps. The hierarchy persists for weaker noise and other evolution times. As the noise weakens, the memory depth of the same reservoirs grows far more slowly than that of an idle qubit, which grows as $1/(1-q)$, and it saturates in the noiseless limit [Fig.~\ref{figS:sim}(b)]. There, the reinitialization of the input register is the only source of forgetting: every time step discards the part of the stored information that the dynamics has moved into the input register. Even a noiseless reservoir therefore forgets, within 75 to 539 time steps for our Ising reservoirs, 175 on average, and within 13 time steps for the Haar-random ones at $n=6$. For the Ising reservoirs, the mean memory depth lies between 16 and 23 time steps for all evolution times from $J\Delta t=1$ to 30 [Fig.~\ref{figS:sim}(c)], far below the idle qubit. For very short evolution times, the input hardly leaves the input register, so the reservoir forgets it at the next injection.

\begin{figure}[t]
\centering
\includegraphics{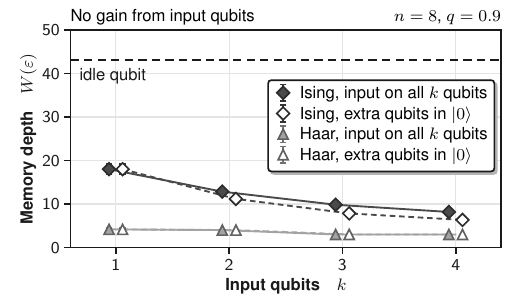}
\caption{\textbf{More input qubits do not help standard reservoirs.} Memory depth of the six Ising and the six Haar reservoirs with $n=8$ versus the number of input qubits, when all input qubits carry the input (filled) and when only the first one does and the others are reinitialized in $\ket0$ (open; mean and standard deviation). For $k=1$, both protocols coincide. Unitary ensembles are sampled separately for each $k$; both protocols share the ensemble at fixed $k$.}
\label{figS:cold}
\end{figure}

\paragraph*{Cold input qubits.} In the runs above, every input qubit carries the input, so a larger $k$ also overwrites more of the reservoir with new input. To separate the two effects, we repeated the runs at $n=8$ for the same reservoirs with only the first input qubit carrying the input and the other $k-1$ input qubits reinitialized in $\ket0$, which makes them pure entropy sinks. The memory depth still falls as $k$ grows [Fig.~\ref{figS:cold}], from 18 time steps for $k=1$ to 11, 8, and 6 for $k=2$, 3, and 4 (Ising, means over the six reservoirs), and from 4 to 4, 3, and 3 (Haar), compared with 13, 10, and 8 (Ising) when all input qubits carry the input. A cold qubit removes entropy, but a generic unitary moves stored information into it as readily as errors, and the next reinitialization erases both. Only dynamics that separates the two, as the error-correcting reservoir does, turns fresh qubits into memory.

\section{Implications for reservoir design, and scope}
\label{sm:scope}

\subsection{Design lessons}

Return to the forecasting task of Fig.~\ref{fig:overview}(a), $y_{t+1}=0.7\,u_t+0.3\,u_{t-100}$, and consider two reservoirs with the same noise, $q=0.9$, and the same $k=2$ input qubits per time step, a reservoir A with $n=10$ qubits and a reservoir B with $n=100$ qubits [Fig.~\ref{fig:limit}(b)]. To forecast $y_{t+1}$, a reservoir must retain $u_{t-100}$, that is, it needs a memory depth of at least 100 time steps at an accuracy that resolves the second term. In fact, it must retain every input of the last 100 time steps at once, a stronger requirement, for which a memory depth of 100 time steps is only a necessary condition. The natural remedy for a reservoir that falls short is to enlarge it. Theorem~\ref{thm:main} shows that this remedy has a limit that does not move with $n$: with two input qubits, the memory depth of every reservoir, A, B, or any larger one, is bounded by the same constant $W_{q,2}(\varepsilon)$, whereas every additional input qubit can multiply the attainable memory depth by a constant factor. Deciding whether 100 time steps lie below $W_{0.9,2}(\varepsilon)$ requires sharper constants (Sec.~\ref{sm:tight}); the theorem identifies the resource, not its numerical setting. The simulations add a practical scale: standard Ising reservoirs remember 16 to 18 time steps at this noise level, for every size from $n=3$ to $n=9$.

\paragraph*{Reservoir size.} A larger reservoir can generate richer features of its recent inputs~\cite{Nakajima2019}, but at fixed $k$ and $q$ its best memory depth is bounded independently of $n$. Large random reservoirs can even suffer from an exponential concentration of their outputs, which makes different inputs hard to distinguish~\cite{Xiong2025,Dao2026}, and in our simulations they forget faster [Fig.~\ref{fig:attain}(c)].

\paragraph*{Input injection.} The number of qubits reinitialized per time step sets the limit. More reinitialized qubits do not guarantee a longer memory, however, because reinitialization also erases whatever those qubits stored, even when the additional qubits are reinitialized in $\ket0$ [Fig.~\ref{figS:cold}]. The limit is reached only if the reservoir dynamics moves the relevant information out of the input register before it is reinitialized and uses the fresh qubits to repair it, as the error-correcting reservoir does.

\paragraph*{Memory requirement of the task.} If a task depends on inputs further back than the memory depth $W(\varepsilon)$, these inputs change every trained readout by at most $2\varepsilon\Norm F$. Estimating how far back the task-relevant inputs lie is therefore a prerequisite for choosing $k$.

\paragraph*{Forecasting horizon.} Predicting many steps ahead does not by itself require a deep memory; what matters is how far back the inputs lie that determine the target. In the task above, even a one-step forecast requires a memory depth of 100 time steps.

\subsection{Relation to limits on noisy quantum computation}
\label{sm:prior}

Theorem~\ref{thm:main} joins a family of results on what noise permits with and without fresh qubits. This section states what each of them bounds and where Theorem~\ref{thm:main} differs.

\paragraph*{What is known and what is new.} That $O(\log T)$ fresh qubits per round suffice to keep a classical bit for $T$ rounds under depolarizing noise follows from the repetition code with majority vote~\cite{vonNeumann1956}, and Sec.~\ref{sm:rep} makes this quantitative. Theorem~\ref{thm:main}(i) is the converse: $k$ fresh qubits per time step buy at most $\exp[O_q(k)]$ time steps of memory, for every number of qubits, every unitary, and every input encoding, whether or not the dynamics has the structure of a code. This does not follow from counting entropy. Noise raises the entropy of every qubit at every time step while only $k$ qubits are reinitialized, so a large reservoir is highly mixed, but a mixed state can store a bit perfectly: $\ketbra00\otimes I/2^{n-1}$ and $\ketbra11\otimes I/2^{n-1}$ are almost completely mixed and perfectly distinguishable (Sec.~\ref{sm:norms}). The proof therefore needs two further ingredients, hypercontractivity, which caps the purity of every reservoir state by a constant independent of $n$ (Lemma~\ref{lem:cap}), and reverse hypercontractivity, which turns the missing purity into forgetting (Lemma~\ref{lem:forget}).

\paragraph*{Without fresh qubits.} Aharonov, Ben-Or, Impagliazzo, and Nisan~\cite{Aharonov1996} showed that noisy reversible circuits, classical or quantum, that receive all their inputs at the start and no fresh bits compute reliably only to a depth logarithmic in their size. The purity recursion reproduces this for depolarizing noise. With $b=0$, Eq.~\eqref{eq:S-washout} gives $\log_2P_t^2\le\theta^tn$, and since
\[
\Norm{\rho-I/d}_1\le\sqrt d\,\Norm{\rho-I/d}_2=\sqrt{d\tr\rho^2-1}=\sqrt{P^2-1},
\]
every state is within trace distance $\varepsilon$ of the completely mixed state after $\ln[n/\log_2(1+4\varepsilon^2)]/(-\ln\theta)=O_q(\log(n/\varepsilon^2))$ time steps, whatever the circuit does. Ben-Or, Gottesman, and Hassidim~\cite{BenOr2013} classified the noise channels for computation without fresh ancillas: depolarizing-type noise permits only a logarithmic time, dephasing a polynomial time, and amplitude damping a time exponential in the number of qubits, because relaxing idle qubits act as a refrigerator that supplies cold ancillas. The cooling reservoir of Sec.~\ref{sm:cooling} is the classical counterpart of this refrigerator, and Sec.~\ref{sm:counter} shows that a bit stored in the computational basis survives pure dephasing forever. Non-unital noise is thus an entropy sink of its own, and noise without a depolarizing component need not erase anything. This is why Theorem~\ref{thm:main} requires a depolarizing component: the assumption cannot be dropped.

\paragraph*{With fresh qubits.} Fawzi, M\"uller-Hermes, and Shayeghi~\cite{Fawzi2022} proved that, at fixed sufficiently small simulation error, a fault-tolerant simulation of a quantum circuit of width $n$ and length $T$ against i.i.d.\ noise given by a non-unitary qubit channel $\mathcal C$ needs at least $\max\{Q(\mathcal C)^{-1}n,\ \alpha_{\mathcal C}\log T\}$ physical qubits, where $Q(\mathcal C)$ is the quantum capacity and $\alpha_{\mathcal C}>0$ depends only on the channel, in a model that allows qubits to be replaced by fresh ones at any time. For unital noise, including depolarizing noise, the bound also permits free and perfect classical computation. This bounds the width of the noisy circuit as a function of its length and leaves open how many of its qubits must be fresh in each round. Theorem~\ref{thm:main} bounds the complementary quantity, the rate of fresh qubits per time step, for an unbounded number of qubits and without classical feedback; the two bounds constrain different resources. The exponential upper bound of Ref.~\cite{Fawzi2022} on the length of fault-tolerant computation under amplitude damping is the quantum counterpart of the exponential memory of the cooling reservoir (Sec.~\ref{sm:cooling}).

Nelson, Rajakumar, and Gullans~\cite{Nelson2026} restrict the gate set instead of the entropy sink: Clifford circuits with resets, in any number and to any states, without intermediate measurements or noiseless classical processing, and interspersed with the depolarizing noise $\cD_q$, map any two input states to within trace distance $\varepsilon$ after $O((1-q)^{-1}\log n\,\log(n/\varepsilon))$ layers. Clifford dynamics therefore cannot store information for longer than a time polylogarithmic in $n$, however many fresh qubits it receives, whereas Theorem~\ref{thm:main} allows every gate set and bounds the memory by $\exp[O_q(k)]$ for every $n$; for a noisy Clifford circuit with $k$ resets per layer, both bounds apply. The two results agree on the construction: the majority vote of Sec.~\ref{sm:rep} is a non-Clifford operation, as long-term memory requires~\cite{Nelson2026}.

\subsection{What the result does not say}

Memory depth is only one requirement of a temporal task. A reservoir must also transform its input history nonlinearly and make the relevant information visible to its readout~\cite{Cao2026}, and the error-correcting reservoir, whose memory scaling is optimal, computes nothing beyond a delayed copy of its input. Theorem~\ref{thm:main} therefore makes no statement about task performance, nonlinear processing capacity, sampling noise, or a quantum advantage. It limits how far back any task can reach but says nothing about performance within that range. In particular, the theorem does not provide an optimal QRC architecture for time series. It does not say how many input qubits a reservoir of 10 qubits should use, which couplings generate the most useful temporal features~\cite{MartinezPena2021,Cao2026}, how much memory a given real-world time series requires, or when a quantum reservoir outperforms a classical one. The simulations cover sizes up to $n=9$; they locate standard reservoirs relative to the limit and do not test the bound, which is a theorem. The result is a design principle; a recipe would also have to answer the questions listed above.

\section{Code and data}
\label{sm:code}

Project repository: \url{https://github.com/eybmits/fresh-qubits-memory}.

The source archive of this Supplemental Material contains the code and data behind every number, table, and figure. The folder \texttt{sim} holds the density-matrix simulator and the memory-depth estimator (\texttt{qrc\_sim.py}), the simulation runs (\texttt{campaign.py}, \texttt{merge.py}), the validation of the code and of the estimator (\texttt{test\_sim.py}, \texttt{validate\_pairs.py}), the error-correcting reservoir (\texttt{rep\_curve.py}), the exact constants, half-lives, washout times, and the cooling reservoir (\texttt{exact\_numbers.py}), and all results as JSON files. The folder \texttt{figures} holds the sources of all figures, and the folder \texttt{checks} independent checks of the constants and of the elementary steps of the proof (\texttt{verify\_constants.py}, \texttt{verify\_simple\_proof.py}). A README file lists the commands that rebuild every figure from these data. Claude Opus 5.5 assisted with code development. Markus Baumann reviewed the code using the checks of Sec.~\ref{sm:numerics} and comparisons with the exact results of Secs.~\ref{sm:washoutlogn} and~\ref{sm:rep}.

\end{document}